\documentclass[ reprint,showpacs,prb,groupedaddress,
 amsmath,amssymb,
 aps,floatfix,
]{revtex4-1}

\usepackage{graphicx}
\usepackage{dcolumn}
\usepackage{braket}
\usepackage[unicode=true,bookmarks=true,bookmarksnumbered=false,bookmarksopen=false,breaklinks=false,pdfborder={0 0 1},backref=false,colorlinks=true]{hyperref}
\hypersetup{ pdfauthor={Le Tuan Anh Ho, Liviu F. Chibotaru}, linkcolor=blue,urlcolor= blue,citecolor= blue,anchorcolor= blue}

\global\long\def\chip{\chi'}
\global\long\def\chipp{\chi''}

\begin{document}


\title{Multiple Relaxation Times in Single-molecule Magnets}

\author{Le Tuan Anh Ho}
\email{anh.holetuan@chem.kuleuven.be}
\affiliation{Theory of Nanomaterials Group, Katholieke Universiteit Leuven, Celestijnenlaan 200F, B-3001 Leuven, Belgium}
\author{Liviu F. Chibotaru}
\email{liviu.chibotaru@kuleuven.be}
\affiliation{Theory of Nanomaterials Group, Katholieke Universiteit Leuven, Celestijnenlaan 200F, B-3001 Leuven, Belgium}

\date{\today}

\begin{abstract}
Multiple relaxation times detected in the ac magnetic susceptibility of several single-molecule magnets have been always assigned to extrinsic factors, such as nonequivalent magnetic centers or effects of intermolecular interactions in the crystal. By solving quantum relaxation equations, we prove that the observed multiple relaxation times can be of intramolecular origin and can show up even in single-ion metal complexes. For the latter a remarkably good description of the coexistent two relaxation times is demonstrated on several experimental examples. This proves the relevance of the intramolecular mechanism of multiple relaxation times in such systems, which is even easier justified in polynuclear magnetic complexes.

\end{abstract}

\maketitle

\maketitle

Single-molecule magnets (SMM) have drawn increasing attention in recent years due to their prospects for storing information and spintronics devices at molecular level \cite{Vincent2012,*Bogani2008,*Leuenberger2001,*Gatteschi2003,*Sessoli1993}. To be an adequate material for these purposes, a SMM should have a large relaxation time of magnetization $\tau$. The latter is routinely extracted from ac magnetic susceptibility measurements. According to the generalized Debye model \cite{casimir1938a,*mcconnell1980,*cole1941}, the relaxation time $\tau$ is associated with the inverse of the frequency at which the out-of-phase susceptibility $\chipp (\omega )$ attains its maximum, $\tau=1/\omega_{\mathrm{max}}$. However, this interpretation of the ac susceptibility becomes confusing with the emergence of lanthanide-based SMMs \cite{Ishikawa2003,*Sessoli2009,*Woodruff2013,*AlDamen2009}, where more and more observations of two maxima in  $\chipp (\omega )$ are reported in polynuclear compounds \cite{Blagg2013b,Hewitt2010,*Lin2009,*Guo2011c,*Amjad2016,*Guo2011a,*Guo2010,*Hewitt2009}. This phenomenon is usually explained by associating each relaxation time to distinct relaxation pathway at magnetic centers of different kinds in these complexes \cite{Blagg2013b,*Hewitt2010,*Lin2009,*Guo2011c,*Amjad2016,*Guo2011a,*Guo2010,*Hewitt2009}. 

Recently observations of a second maximum in $\chipp (\omega )$ have been also reported for mononuclear SMMs \cite{Rinehart2010,Miklovic2015,Holmberg2015,Gupta2016,Jeletic2011,Habib2013,Habib2015,Ruiz2012,Lucaccini2016b,Gregson2015,Li2016,Cosquer2013}. This by all means cannot be rationalized by the previous argument and thus raises a question on the mechanism behind the existence of the second relaxation time \footnote{The studies of diluted U(H$_2$BPz$_2$)$_3$ have shown a relationship between the existence of the secondary relaxation 
process and the degree of dilution, pointing on the importance of intermolecular interaction for its observation \cite{Meihaus2011,*Meihaus2015}}, with possible implications not only for mononuclear but also polynuclear SMMs. Answering this question will certainly advance our understanding of relaxation processes in magnetic molecules and will contribute to an adequate interpretation of ac susceptibility data in such systems. On a practical side, the deep knowledge of the details of relaxation in magnetic complexes is indispensable for a rational design of efficient SMMs. In this work, we prove that the observed secondary relaxation process 
in mononuclear SMMs has an intramolecular origin. The derived analytical expressions display the conditions for the observability of the second relaxtion time in the ac magnetic susceptibility. The mechanism underlying this effect is generic and may be relevant for strongly exchange-coupled polynuclear SMMs as well. 

\begin{figure}[h]
\centering{}\includegraphics[scale=0.4]{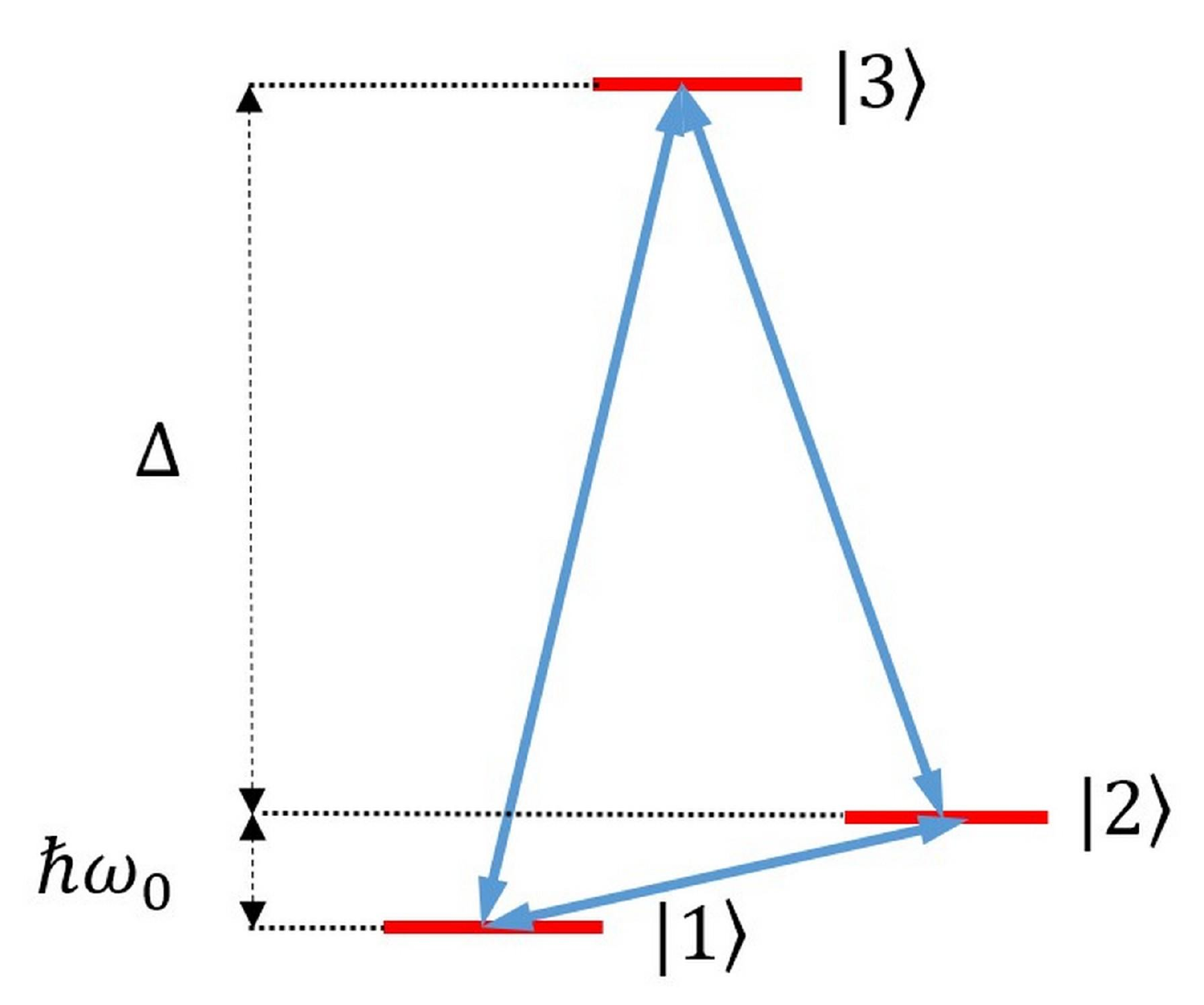}\caption{Electronic structure of the model\label{fig:model}}
\end{figure}

A system with $n$ electronic states in a thermal bath could relax via several relaxation modes with the rates $\lambda_{i}$, $i=1,\ldots n$, corresponding to the eigenvalues of the relaxation rate matrix \cite{Gatteschi2006,garanin2008,Leuenberger2000}. One of these eigenvalues, $\lambda_{1}$, is zero, corresponding to thermodynamic equilibrium. Having this in mind, a minimal model which could be considered is the one involving three electronic levels (Fig. \ref{fig:model}). In this model the external dc magnetic field is directed along the easy magnetic axis of the ground quasi doublet. The corresponding states of the quasi doublet,
$\ket{1}$ and $\ket{2}$, are separated by a relatively small gap $\omega_{0}$, whereas the third state $\ket{3}$ is supposed to lie at $\Delta\gg\hbar\omega_{0}$. To measure ac susceptibility, we apply an additional small ac magnetic field $h\left(e^{i\omega t}+e^{-i\omega t}\right)$ along the main magnetic axis of the quasi doublet $z$. The Hamiltonian of the system is then $\mathcal{H}=\mathcal{H}_{0}+V$. Here $V=-m_{z}h_{z}(t)=-\sum_{i,j}m_{ij}\ket{i}\bra{j}h\left(e^{i\omega t}+e^{-i\omega t}\right)$ is the ac component of the Zeeman Hamiltonian and $m_{ij}$ are matrix elements of the magnetic momentum $m_{z}$ on the states of the model. 

Following the experimental conditions, we consider further the temperature domain $\hbar\omega_{0}\ll kT$, which allows us to approximate the relaxation rate matrix as  
\begin{equation}
\Phi\approx\Gamma_{21}\left(\begin{array}{ccc}
-\left(1+\alpha\right) & 1 & c\alpha\\
1 & -\left(1+\alpha\right) & c\alpha\\
\alpha & \alpha & -2c\alpha
\end{array}\right),\label{eq:reduce relaxation matrix}
\end{equation}
where $\alpha\equiv\Gamma_{31}/\Gamma_{21}$ and $c\equiv\Gamma_{13}/\Gamma_{31}=\exp\left(\Delta/kT\right)$. 

The electronic levels in Fig. 1 are eigenvalues of $\mathcal{H}_{0}$ including the effect of the dc magnetic field and, therefore, are all magnetic ($m_{ii}\neq 0$). They basically arise from the Zeeman splitting of the ground and the first excited (quasi) doublets (for the latter only the lowest Zeeman component $\ket{3}$ is shown in Fig. 1). It is assumed, in line with experiments where secondary relaxation proccess was observed, that the applied field (several tenths of Tesla) is sufficient for suppression of tunneling in both these (quasi) doublets. At the same time the nature of the metal ion is not important, so the further consideration equally applies to Kramers and non-Kramers ions. 

The typical situation for low-temperature relaxation is $kT\ll \Delta $, when the population of the highest state $\ket{3}$ is small and its variation during the relaxation can be neglected \cite{abragam1970}. In such case the relaxation basically occurs within the lowest two states with the rate  $\Gamma =\Gamma_{\mathrm{Orbach}}+\Gamma_{\mathrm{direct}}+\Gamma_{\mathrm{Raman}}$ \footnote{Generally, a tunneling relaxation rate, $\Gamma_{\mathrm{tunneling}}$, should be added to the right hand side of this expression, however, it is suppressed in the present case}, where the Orbach relaxation rate, $\Gamma_{\mathrm{Orbach}}=\Gamma_{0}\exp\left[U_{\mathrm{eff}}/kT\right]$, includes the effect of two direct relaxation rates via the exited state $\ket{3}$ \cite{abragam1970}. In our treatment, however, $kT$ will be considered of the order of $\Delta$ (vide infra), in which case the constant $c$ in Eq. (\ref{eq:reduce relaxation matrix}) is not expected to be large. Then this situation should be treated via the solution for the full density matrix involving the three states. Denoting by $\delta\boldsymbol{\rho}$ its deviation from the equilibrium value (${\boldsymbol{\rho}}^{\text{eq}}$) induced by the ac magnetic field, the equation for the diagonal components has the form:
\begin{equation}
\frac{d}{dt}\delta\boldsymbol{\rho}=\Phi\delta\boldsymbol{\rho}+\mathbf{f^{\pm}},\label{eq:vectorized linear response equation}
\end{equation}
where $\mathbf{f^{+}}$ and $\mathbf{f^{-}}$ are defined from the relation $f_{\alpha}^{+}e^{i\omega t}+f_{\alpha}^{-}e^{-i\omega t}\equiv\rho_{\alpha\alpha}^{\mathrm{eq}}\sum_{\beta\ne\alpha}\Gamma_{\beta\alpha}\left(V_{\beta\beta}-V_{\alpha\alpha}\right)/kT$.
Assuming $\mathbf{f^{+}=f^{-}\equiv f_{0}}$ as is usually the case, the solution
of Eq. \eqref{eq:vectorized linear response equation} is obtained in the form:
\begin{align}
\delta\boldsymbol{\rho} & =2\sum_{\mu\ne1}\left(\frac{\lambda_{\mu}}{\lambda_{\mu}^{2}+\omega^{2}}\cos\omega t+\frac{\omega}{\lambda_{\mu}^{2}+\omega^{2}}\sin\omega t\right)\frac{\left(\mathbf{L}{}_{\mu}\ldotp\mathbf{f_{0}}\right)}{\left(\mathbf{L}_{\mu}\ldotp\mathbf{R}_{\mu}\right)}\mathbf{R}_{\mu}\label{eq:general linear response}
\end{align}
where $\lambda_{\mu},\mathbf{L}_{\mu},$ and $\mathbf{R}_{\mu}$ are respectively eigenvalues, left, and right eigenvectors of the relaxation rate matrix $\Phi$ \cite{garanin2008,a1}. The off-diagonal elements of the density matrix are found from the equations: 
\begin{equation}
\frac{d}{dt}\delta\rho_{\alpha\beta}=iV_{\alpha\beta}\left(\rho_{\alpha\alpha}^{\mathrm{eq}}-\rho_{\beta\beta}^{\mathrm{eq}}\right)-\left(i\omega_{\alpha\beta}+\gamma_{\alpha\beta}\right)\delta\rho_{\alpha\beta}.
\end{equation}

As evidenced from above equations, under the assumption $\hbar\omega_{0}\ll kT$ the oscillations of the off-diagonal elements of the density matrix $\delta\rho_{12}$ and $\delta\rho_{21}$ are very small and can be neglected in comparison to the diagonal elements. Moreover, in a nonzero magnetic field the off-diagonal matrix elements of the easy-axis magnetic moment matrix of a SMM ($m_{\alpha\beta}$) are often much smaller than the diagonal ones $(m_{\alpha\alpha})$. Hence, the effect of the off-diagonal magnetic moment matrix elements can be safely ignored in the calculations of the linear response, validating the relation $\mathrm{Tr}\left(\hat{m}\cdot\delta\hat{\rho}\right)\approx\mathbf{m}\cdot\delta\boldsymbol{\rho}$. Having this in mind and making use of the relation $h\left(\chip\cos\omega t+\chipp\sin\omega t\right)=\mathrm{Tr}\left(\hat{m}\cdot\delta\hat{\rho}\right)$, we obtain the following expressions for the ac susceptibility after diagonalizing the relaxation rate matrix $\Phi$:%
\begin{align}
\chip & =\frac{1}{T}\frac{c}{1+2c}\left[m_{11}^{2}\frac{1}{1+\omega^{2}\tau_{2}^{2}}+\frac{1}{1+2c}m_{33}^{2}\frac{1}{1+\omega^{2}\tau_{3}^{2}}\right],\label{eq:in-phase ac susceptibility}\\
\chipp & =\frac{1}{T}\frac{c}{1+2c}\left[m_{11}^{2}\frac{\omega\tau_{2}}{1+\omega^{2}\tau_{2}^{2}}+\frac{1}{1+2c}m_{33}^{2}\frac{\omega\tau_{3}}{1+\omega^{2}\tau_{3}^{2}}\right],\label{eq:out-of-phase ac susceptibility} \end{align} 
where $\tau_{2}^{-1}=\lambda_{2}=\Gamma_{21}\left(2+\alpha\right)=2\Gamma_{21}+\Gamma_{31}$ and $\tau_{3}^{-1}=\lambda_{3}=\Gamma_{31}\left(1+2c\right)$. Note that the rate of relaxation between the two lowest levels, $\Gamma_{21}+\Gamma_{12}\approx2\Gamma_{21}$, includes the direct and the Raman	processes \cite{Note2}, while $\Gamma_{31}$ is the relaxation rate of the Orbach process, $\Gamma_{\mathrm{Orbach}}$ \footnote{Given the explicit involvement of the state $\ket{3}$ in the present description of relaxation, Eq. (\ref{eq:general linear response}), $\Gamma_{31}$ refers to a direct relaxation process. However, we still call it $\Gamma_{\mathrm{Orbach}}$ for the sake of convenience}. Then we recover for one relaxation rate the familiar expression, $\tau_{2}^{-1}=\Gamma_{\mathrm{direct}}+\Gamma_{\mathrm{Raman}}+\Gamma_{\mathrm{Orbach}}$ \cite{Note2}, while the other is rewritten as $\tau_{3}^{-1}=\Gamma_{\mathrm{Orbach}}\left(1+2c\right)$. 

Eqs. \eqref{eq:in-phase ac susceptibility} and \eqref{eq:out-of-phase ac susceptibility}, look as sums of two Debye functions, often used for the phenomenological description of the ac susceptibility data displaying two relaxation times \cite{Guo2011b}. The important difference is the restriction on the ratio of these two Debye function contained in Eqs. \eqref{eq:in-phase ac susceptibility} and \eqref{eq:out-of-phase ac susceptibility}. It will be shown that this very distinction is the major reason behind the difficulty of the observation of the secondary relaxation process in ac susceptibility experiments.

As can be easily seen, when the ratio $\kappa\equiv\left(m_{33}/m_{11}\right)^{2}/\left(1+2c\right)$ is negligible, $\chipp\left(\omega\right)$ and the Cole-Cole plot \cite{Gatteschi2006} have only one maximum at the frequency $\omega_{\max}=\tau_{2}^{-1}$. This implies that the relaxation rate extracted from ac susceptibility measurements is indeed a simple sum of the rates from individual relaxation processes (see above). Hence, the present proof is a justification for the wide use of the formula $\Gamma =\Gamma_{\mathrm{Orbach}}+\Gamma_{\mathrm{direct}}+\Gamma_{\mathrm{tunneling}}+\Gamma_{\mathrm{Raman}}$ for the interpretation of measured relaxation rates.

Remarkably, when $\kappa$ is of the order of unity, a second maximum in $\chipp\left(\omega\right)$ and Cole-Cole plot arises. Whereas the first maximum corresponds to the familiar relaxation rate $\tau_{2}$ mentioned above, the second maximum corresponds to $\tau_{3}$, which depends solely on $\Gamma_{\mathrm{Orbach}}$. That is, the nature of the secondary relaxation process is entirely related to the excited state $\ket{3}$. 

Note that although the observation of the secondary relaxation process requires both relaxation rates $\lambda_{2}$ and $\lambda_{3}$ not exceeding the limiting frequency of ac susceptibility measurements, the  existence of two maxima in $\chipp\left(\omega\right)$ and Cole-Cole plot does not depend explicitly on their relative values.  As a consequence, keeping only the smallest (nonzero) relaxation rate while neglecting the larger ones, as it is usually done in the simulation of recovery magnetization measurements,  may lead to a wrong analysis of the ac susceptibility data.

Trying to find favorable conditions for the observation of two peaks in $\chipp\left(\omega\right)$, one could first think of increasing the temperature in order to reduce $\kappa$ to the order of unity. This is not always a practical solution, because increasing the temperature results in larger relaxation rates $\lambda_{2}$ and $\lambda_{3}$, which eventually might leave the available frequency domain of ac susceptibility measurements (usually < $1500$ Hz). Another strategy would be the design of systems with a low excitation energy $\Delta$. This is not practical either because lowering this energy gap enhances the Orbach relaxation rate which automatically increases $\lambda_{2}$ and $\lambda_{3}$. One can infer from this analysis that the conditions of observability of two maxima in $\chipp\left(\omega\right)$ are hardly met, which is fully supported by the experimental situation.

To facilitate further analysis, we consider a temperature domain where the direct relaxation rate $\Gamma_{\mathrm{direct}}\approx aT$ \cite{abragam1970} dominates the Raman relaxation rate. Together with $\Gamma_{\mathrm{Orbach}}=\Gamma_{0}/(\exp\left[\Delta/kT\right]-1)$, the relaxation rate eigenvalues entering Eq. (\ref{eq:out-of-phase ac susceptibility}) are now of the form 
\begin{align}
\lambda_{2} & =aT+\Gamma_{0}/\left(c-1\right),\label{eq:lambda2}\\
\lambda_{3} & =\Gamma_{0}\left(2c+1\right)/\left(c-1\right).\label{eq:lambda3}
\end{align}

\begin{figure}
\begin{tabular}{ll}
{\footnotesize{}(a)} & {\footnotesize{}(b)}\tabularnewline
\includegraphics[scale=0.3]{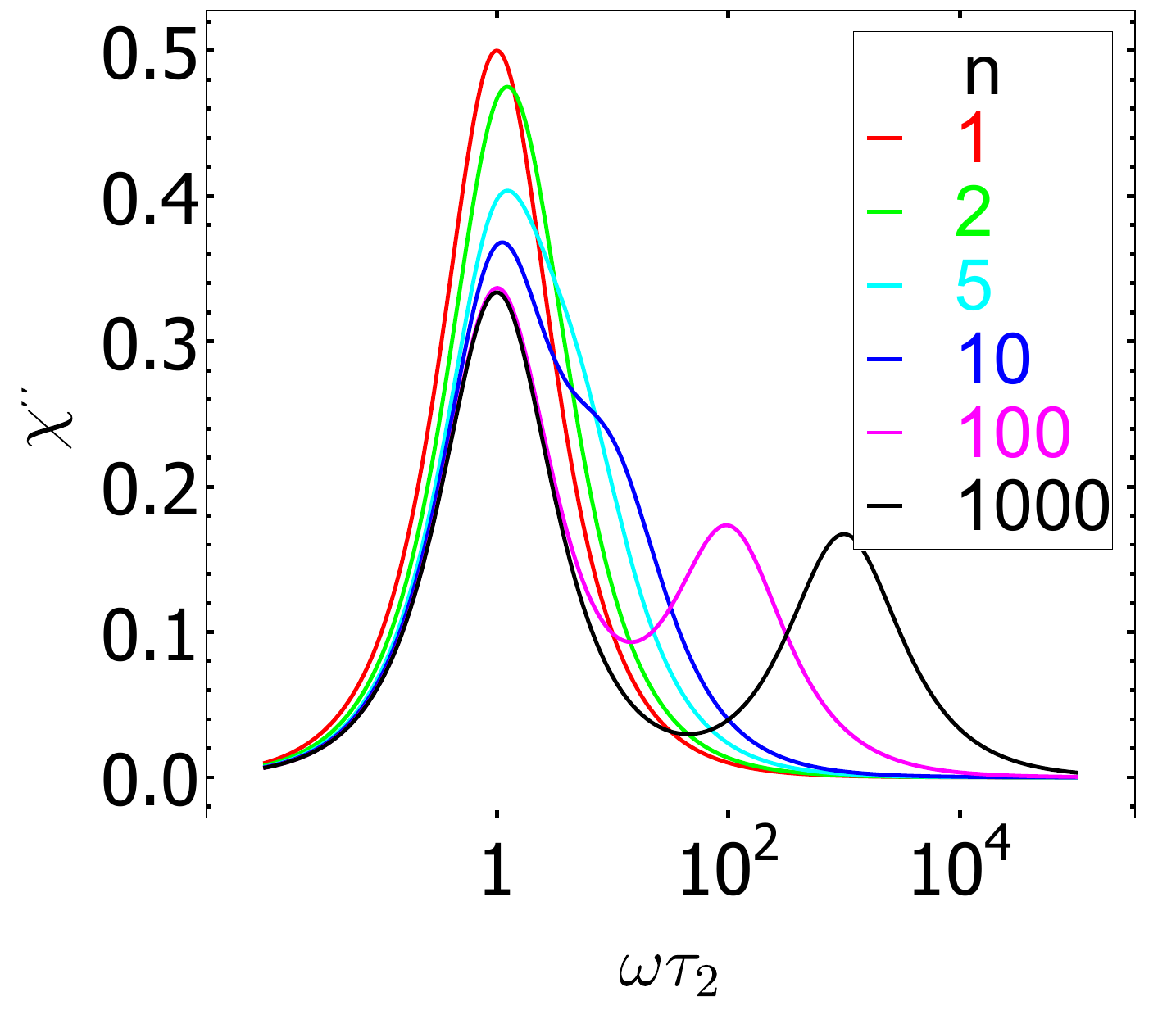} & \includegraphics[scale=0.3]{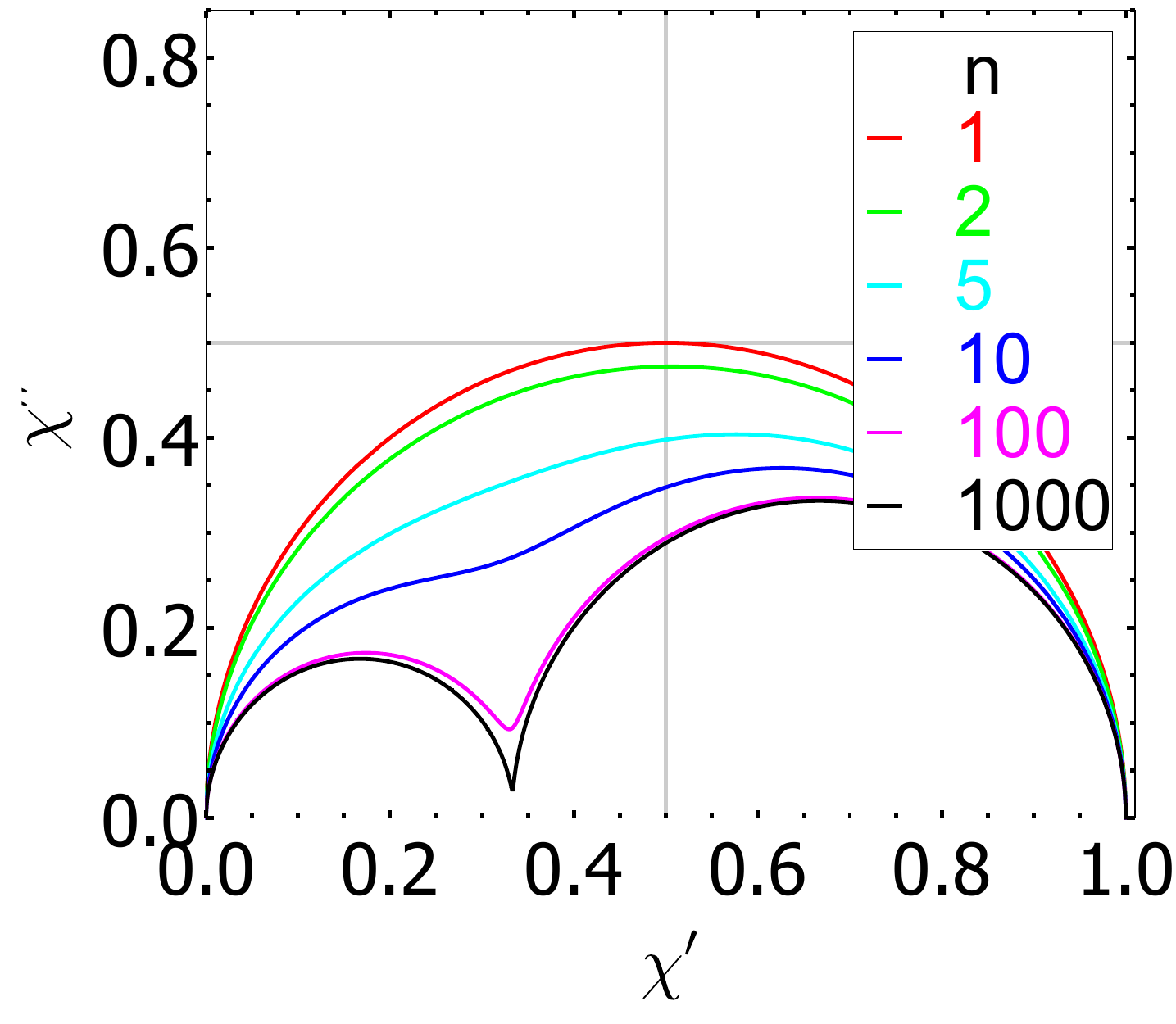}\tabularnewline
\end{tabular}\caption{$\chipp\left(\omega\right)$ and Cole-Cole plots for various values of parameter $n=\tau_{2}/\tau_{3}$ and $\kappa$ = 0.5. A normalization factor $1/\left(1+\kappa\right)$ was used in order to bring the susceptibility to a conventional domain $\chip\in\left(0,1\right]$. 
\label{fig:n-vary}}
\end{figure}

Besides difficulties mentioned above, there is an additional factor making the observation of the secondary relaxation process hard.
Figure \ref{fig:n-vary} shows the frequency dependence of $\chipp$ and the Cole-Cole plot for different values of the ratio $n=\tau_{2}/\tau_{3}$ at a fixed value of $\kappa$. We can see that the second peak in $\chipp\left(\omega\right)$ only appears when $n$ is large enough to separate two peaks from each other. The critical value of $n$ for the arising of the second peak depends on the value of $\kappa$. As for the location of two peaks, they are found at $\omega\tau_{2,3}\approx1$ as expected. For intermediate values of $n$, a shoulder is seen indicating a transition from single-maximum to two-maxima regime. 

An instructive conclusion can be drawn from the analysis of the Cole-Cole plots in Fig. \ref{fig:n-vary}b. Even when the criterion for the existence of two maxima is not fulfilled there is a marked deviation from semicircle shape in these plots with increasing of $n$. These deviations were always interpreted as originating from a distribution of relaxation times among the SMMs in a crystal \cite{Gatteschi2006}. We see that such interpretations can be misleading due to the closeness of a secondary relaxation process.

\begin{figure}
\begin{tabular}[t]{ll}
{\footnotesize{}(a)} & {\footnotesize{}(b)}\tabularnewline
\includegraphics[scale=0.3]{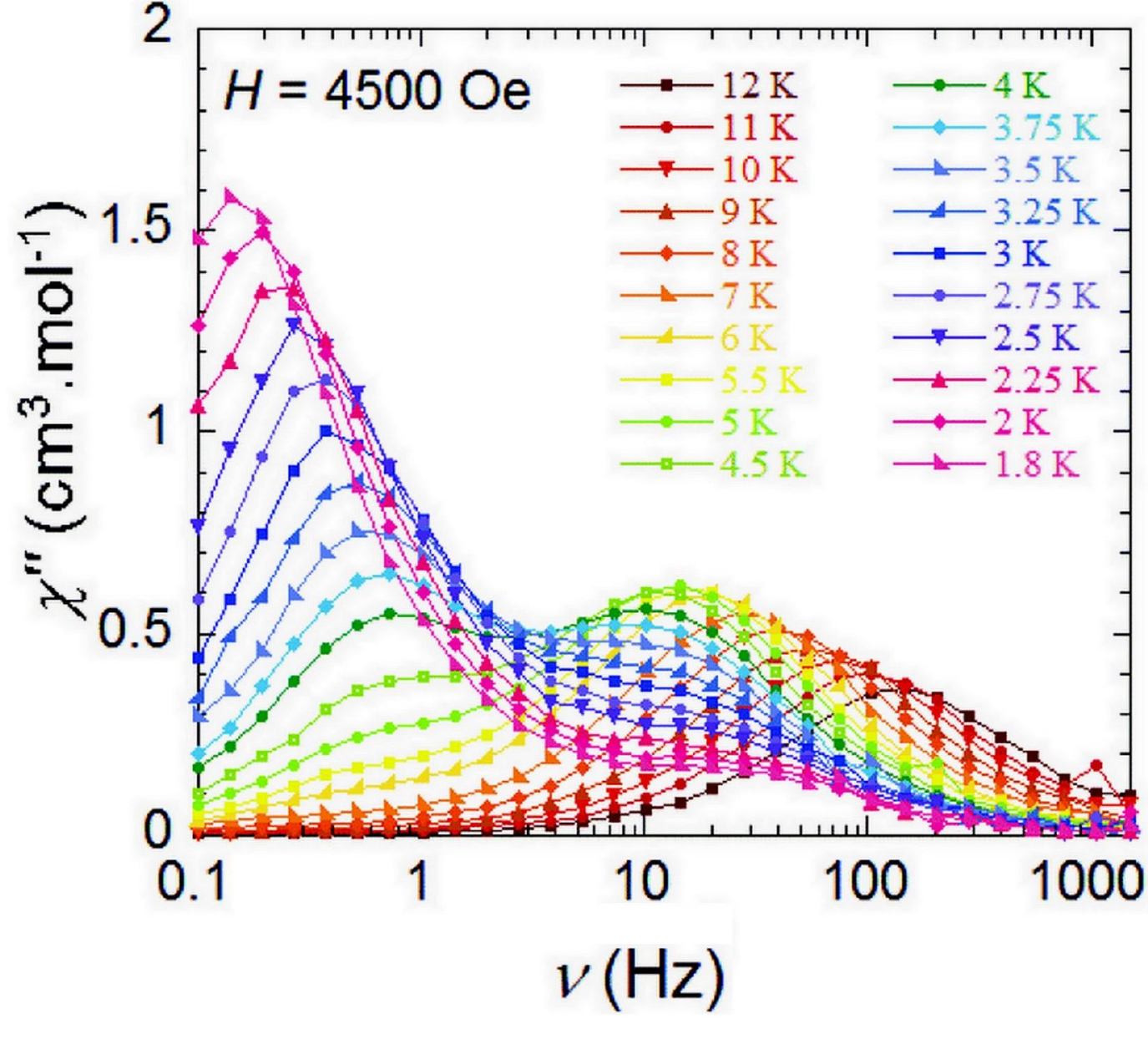} & \includegraphics[scale=0.3]{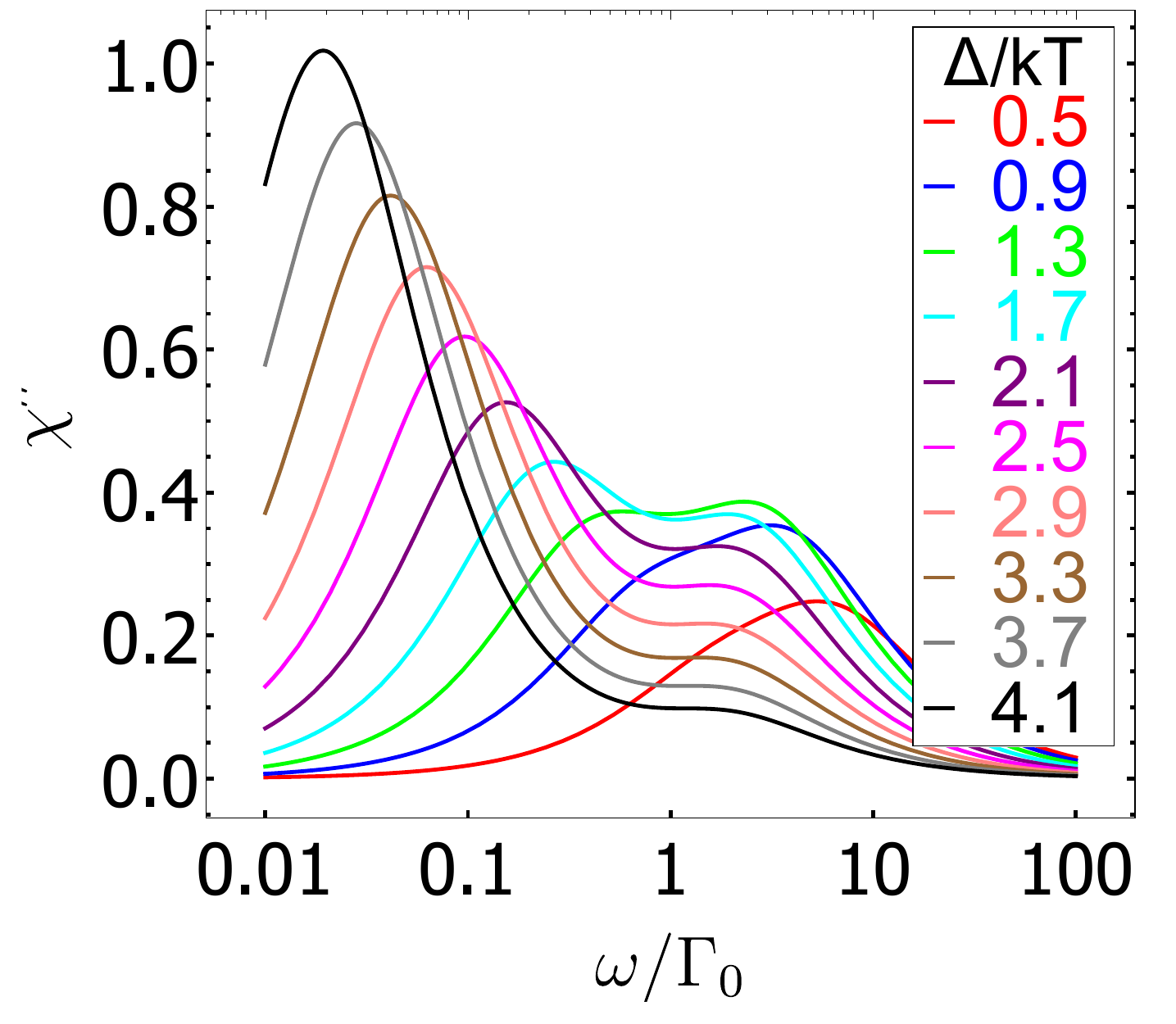}\tabularnewline
\multicolumn{2}{c}{(c)}\tabularnewline
\multicolumn{2}{c}{\includegraphics[scale=0.3]{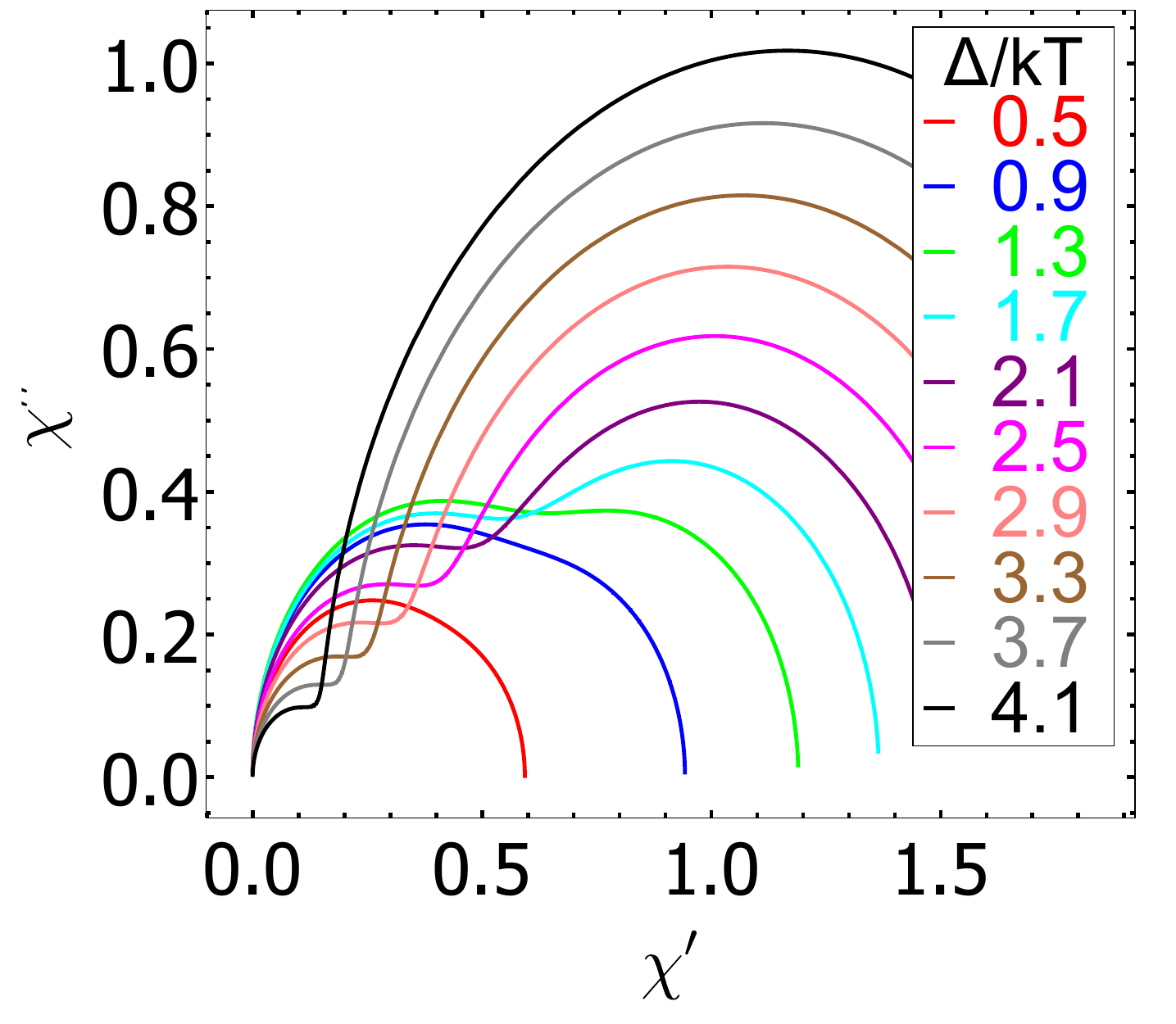}}\tabularnewline
\end{tabular}\caption{(a) Frequency dependence of out-of-phase susceptibility for Gd-EDTA. From Holmberg \textit{et al.} \cite{Holmberg2015} - reproduced by permission of The Royal Society of Chemistry. (b) \& (c) Out-of-phase susceptibility and Cole-Cole plot from our model with parameter $A=0.01$, $m_{33}/m_{11}=3$, and $\Delta/kT\in\left[0.5,4.1\right]$. \label{fig:Holmberg}}
\end{figure}

To investigate the temperature dependence in different relaxation regimes, we introduce dimensionless variables  $\Delta/kT$ and $\Lambda_{i}\equiv\lambda_{i}/\Gamma_{0}$. Thus we have $\Lambda_{2}=A\left(\Delta/kT\right)^{-1}+1/(c-1)$ and $\Lambda_{3}=\left(2c+1\right)/\left(c-1\right)$, where $A\equiv a\Delta/k\Gamma_{0}$ characterizes the relative strength of direct and Orbach processes. Figure \ref{fig:Holmberg}a shows a qualitative comparison of the out-of-phase susceptibility derived from our model with the experimental data from Holmberg \textit{et al.} \cite{Holmberg2015}. From our calculations, this behavior is typical for all values of $A<1$. A Cole-Cole plot is also shown in Figure \ref{fig:Holmberg}c. As follows from the figure, when $T$ decreases, the transition point in Cole-Cole plot shifts from right to left. At the same time, the right hand side semicircle grows and the left hand side one shrinks. Physically, this means that the slow relaxation mode associated with $\tau_{2}$ is getting more and more influential. Apart from this behavior, the rightmost point of $\chipp$ also has the tendency to move rightwards which can be explained by the dominant effect of the factor $\Delta/kT$ on the value of $\chip_{\mathrm{rightmost}}\propto\left(\Delta/kT\right)c/\left(1+2c\right)\times\left[1+\left(m_{33}/m_{11}\right)^{2}/\left(1+2c\right)\right]$. Remarkably, besides the data of Holmberg \textit{et al.} \cite{Holmberg2015}, this kind of behavior is also found in good qualitative agreement with the experimental ac susceptibility of SMMs showing a secondary relaxation process in non-zero field given by Rinehart \textit{et al.} \cite{Rinehart2010} (see Figs. S10-S13 therein), Jeletic \textit{et al.} \cite{Jeletic2011} (see Fig. 3 for 200 Oe and 600 Oe therein), and Habib \textit{et al.} \cite{Habib2013} (see Fig. S16 therein). 

\begin{figure}
\begin{tabular}[t]{ll}
{\footnotesize{}(a)} & {\footnotesize{}(b)}\tabularnewline
 \includegraphics[scale=0.3]{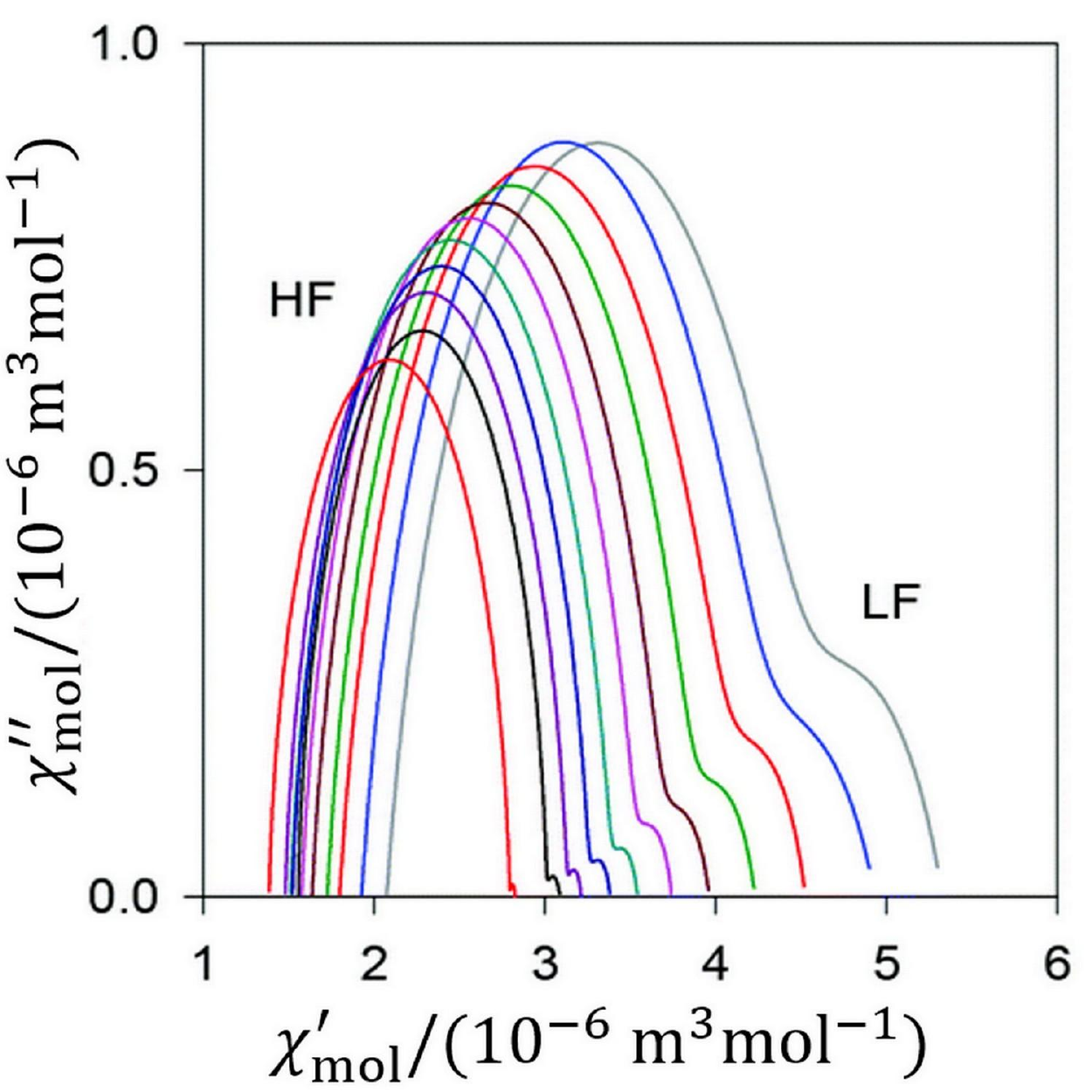} & \includegraphics[scale=0.3]{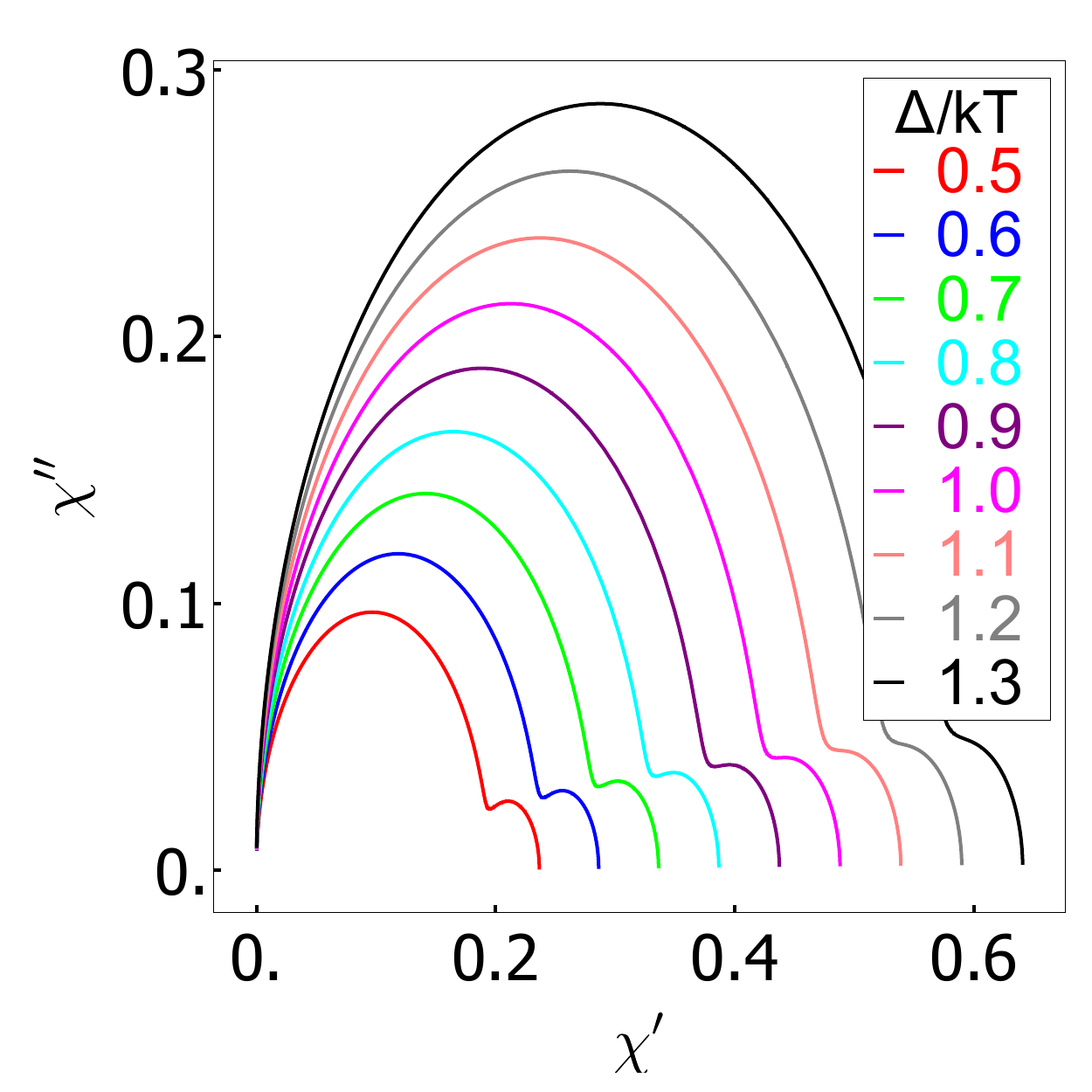}\tabularnewline
{\footnotesize{}(c)} & {\footnotesize{}(d)}\tabularnewline
\includegraphics[scale=0.3]{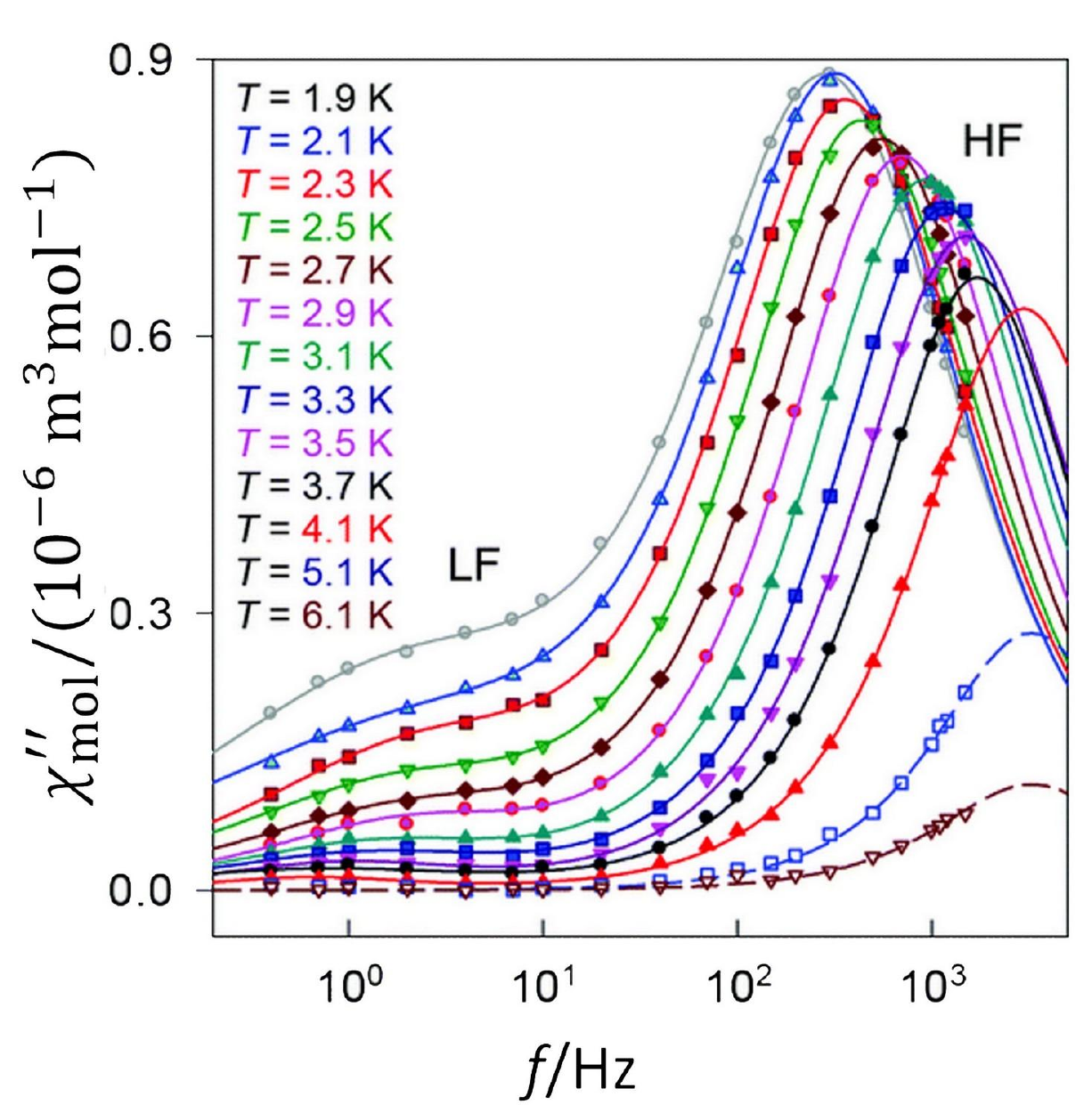} & \includegraphics[scale=0.3]{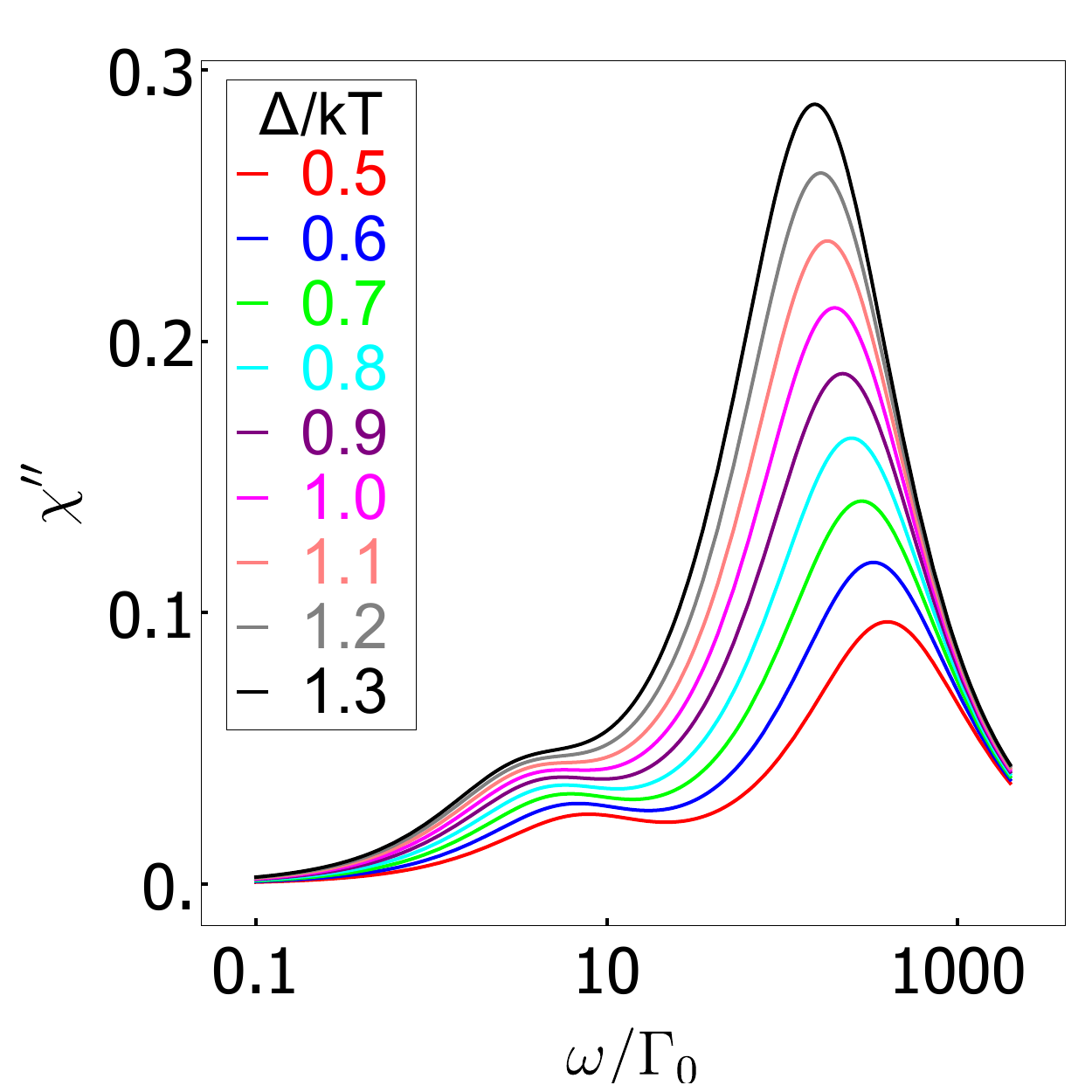}\tabularnewline
\end{tabular}
\caption{(a) Cole-Cole plot and (c) out-of-phase susceptibility for {[}Ni(pydc)(pydm){]}.H$_2$O under applied magnetic field of 0.2 T. Adapted from Miklovi\v{c} \textit{et al.} \cite{Miklovic2015} with permission from The Royal Society of Chemistry. (b) \& (d) Similar plots from the present model for the parameters $A=200$, $m_{33}/m_{11}=1$, and $\Delta/kT\in\left[0.5,1.3\right]$.
\label{fig:Miklovic}} 
\end{figure}

\begin{figure}
\centering{}
\begin{tabular}{ll}
{\footnotesize{}(a)} & {\footnotesize{}(b)}\tabularnewline
\includegraphics[scale=0.3]{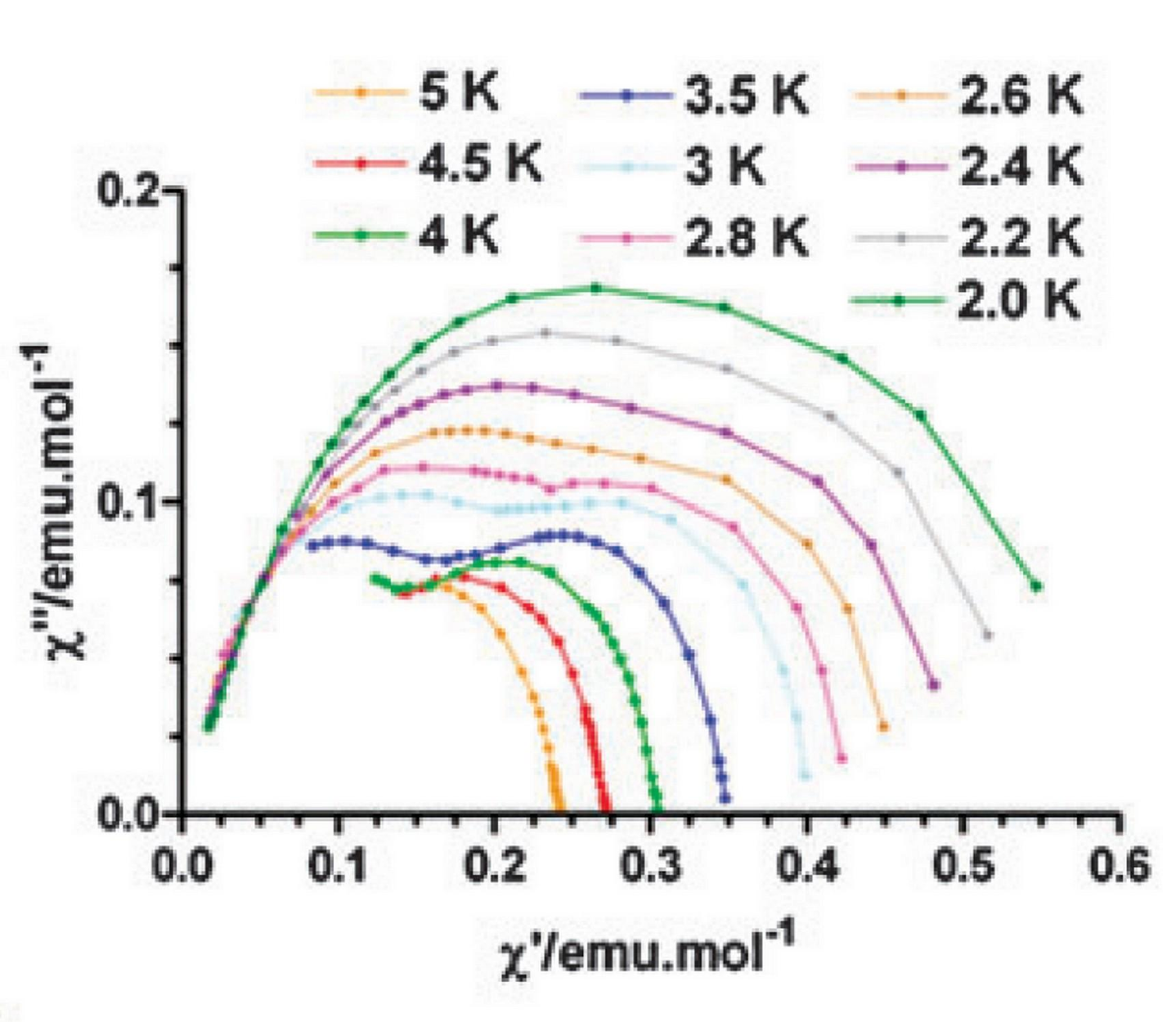} & \includegraphics[scale=0.3]{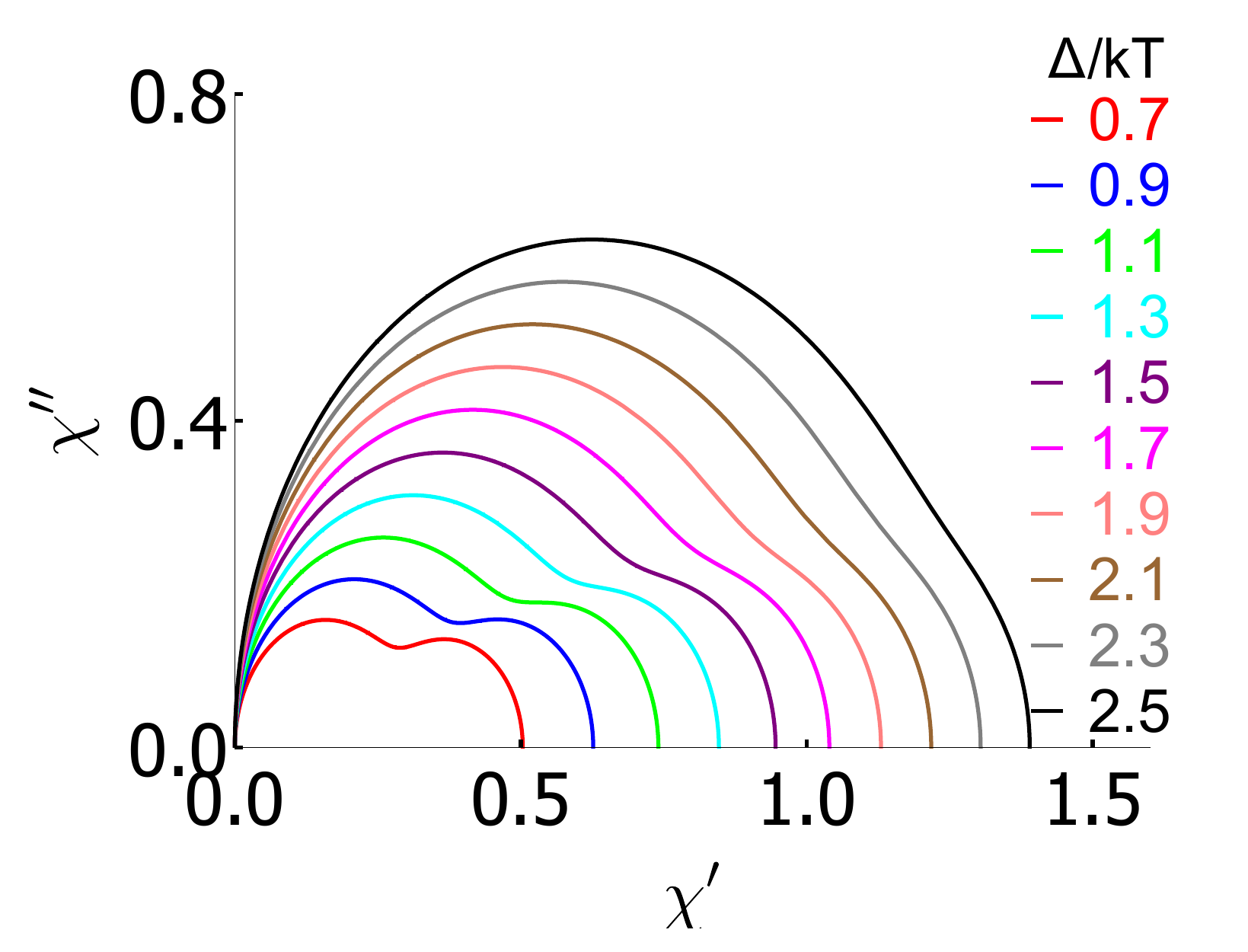}\tabularnewline
\end{tabular}\caption{(Left) Cole-Cole plot for the diluted {[}Dy(H$_2$L)(NO$_3$)$_3${]}.2CH$_3$OH under applied magnetic field of 1000 Oe. Reproduced from Ruiz \textit{et al.} \cite{Ruiz2012} with permission from The Royal Society of Chemistry. (Right) A similar plot from the present model for the parameters $A=50$, $m_{33}/m_{11}=2$, and $\Delta/kT\in\left[0.7,2.5\right]$.
\label{fig:Ruiz}}
\end{figure}

Figures \ref{fig:Miklovic} and \ref{fig:Ruiz} show  comparisons of the present theory with recent experimental data from Miklovi\v{c} \textit{et al.} \cite{Miklovic2015} and Ruiz \textit{et al.} \cite{Ruiz2012}, respectively, which are typical for ac susceptibility behavior in the case $A>1$. In contrast to the previous case, with decrease of $T$ the transition point in Cole-Cole plot shifts from left to right. In the meanwhile, the effect of the fast relaxation process, which is now associated with $\tau_{2}$ instead of $\tau_{3}$, gradually dominates till two processes merge into one. This behavior is abnormal in the sense that the relaxation rates extracted from ac susceptibility measurements are different from the ones measured in recovery magnetization experiments, in which the slowest relaxation rate is derived. It is interesting to note that in both cases, the relaxation rate extracted from ac susceptibility measurement is always reproduced by the conventional expression for the relaxation rate, i.e. a sum of direct, Raman and Orbach relaxation rates. This kind of behavior derived from the present model for $A>1$ also agrees with the report from Li \textit{et al.} \cite{Li2016} (see Fig. 3b therein).

To summarize, there are three reasons explaining why two maxima of ac susceptibility are not often observed in mononuclear SMMs. First, one of the two relaxation rates, $\lambda_2$ or $\lambda_3$, is to high to be detected by conventional ac susceptibility measuring setup. Second, the weight of the strength of the secondary relaxation process ($\kappa$) is too small. Third, the two maxima are too close to each other resulting in their overlap. This third reason might be responsible for the strong dependence of the existence of the second peak on the applied dc field and dilution \cite{a1}, which has been found in the works of Jeletic \textit{et al.} \cite{Jeletic2011} and Habib \textit{et al.} \cite{Habib2013,Habib2015}. 

The results obtained here for a three-level model, allow to rationalize some general features of ac susceptibility in a multi-level systems. First, they explain why it is still difficult to detect more than one maximum of $\chipp$ in these more general systems too. Despite the fact that in a system with $n$ states the relaxation rate matrix can have $n-1$ nonzero eigenvalues and, accordingly, the out-of-phase susceptibility might have up to $n-1$ maxima, the difference in the order of magnitude between these eigenvalues is often huge. Consequently, even if the factor $\kappa_{i}$ corresponding to each theoretical maximum $i^{\mathrm{th}}|_{i>2}$ could be of the order of unity, the frequency $\omega_{\max}^{\left(i\right)}$ at which it should be placed in  $\chipp (\omega )$ is likely to be outside the available range for ac susceptibility measurements. 

A second observation concerns the individual relaxation processes between the two states of the ground (quasi) doublet: direct, tunneling, and Raman. As we already mentioned, it is widely accepted and confirmed in the present work that one of the eigenvalues of the relaxation rate matrix equals to the sum of the relaxation rates of all these individual processes, $\lambda_{2}=\Gamma_{\mathrm{Orbach}}+\Gamma_{\mathrm{direct}}+\Gamma_{\mathrm{Raman}}+\Gamma_{\mathrm{tunneling}}$, where $\Gamma_{\mathrm{direct}}+\Gamma_{\mathrm{tunneling}}+\Gamma_{\mathrm{Raman}}=\Gamma_{21}+\Gamma_{12}$. However, from the general form of the relaxation rate matrix $\Phi$, we have $-\mathrm{Tr}\left(\Phi\right)=\Gamma_{12}+\Gamma_{21}+\Gamma_{31}+\ldots=\lambda_{2}+\sum_{\mu>2}\lambda_{\mu}$. This shows that $\lambda_{2}$ has absorbed the direct, tunneling and Raman relaxation processes
between the two ground states. As a result, these relaxation processes will not contribute to other $\lambda_{i}$ and play no role in the corresponding maxima of out-of-phase susceptibility. 

Despite the fact that the model devised here was applied for simulations of ac susceptibility in single crystals of SMMs molecules, all derived conclusions remain unchanged for powder samples. In fact, although there is a distribution of the relaxation rates $\lambda_{2}$ and $\lambda_{3}$ due to different orientations of the external magnetic field with respect to the frames of SMM molecules in a powder sample, this distribution is continuous and its dispersion is expected to be small in comparison to $\left|\lambda_{2}-\lambda_{3}\right|$. As a consequence, there will be two separated groups of relaxation rates eigenvalues concentrating around some average value $\bar{\lambda}_{2}$ and $\bar{\lambda}_{3}$. These two groups are sufficiently far apart from each other, while inside each of the group the eigenvalues are not as different as to allow individual peaks to appear. 

The intramolecular mechanism discussed here could also be responsible for the observation of secondary relaxation process in strongly coupled polynuclear SMMs as well. Indeed, in such kind of systems the strong interaction between the magnetic centers, at sufficiently low temperature, leads to collective relaxation processes not divisible onto individual magnetic sites. Then the only difference from the physical situation discussed here is a larger number of electronic levels which might be involved. 

In summary, we have proposed an intramolecular mechanism for multiple realaxation times observed in SMMs. Via a microscopic treatment of a three-level model, we have proved analytically that in general two maxima in ac susceptibility can occur even in mononuclear SMMs. The physical requirement for that is the existence of several relaxation modes in the system. Despite its simplicity, our theory shows a very good qualitative agreement with most experimental data where the existence of the secondary relaxation process was reported. The conclusion drawn here is general and could be applied, in particular, to the rationalization of similar phenomena in strongly coupled polynuclear SMMs.

Le Tuan Anh Ho would like to acknowledge the financial support from the Flemish Science Foundation (FWO). 


\begin{thebibliography}{43}%
\makeatletter
\providecommand \@ifxundefined [1]{%
 \@ifx{#1\undefined}
}%
\providecommand \@ifnum [1]{%
 \ifnum #1\expandafter \@firstoftwo
 \else \expandafter \@secondoftwo
 \fi
}%
\providecommand \@ifx [1]{%
 \ifx #1\expandafter \@firstoftwo
 \else \expandafter \@secondoftwo
 \fi
}%
\providecommand \natexlab [1]{#1}%
\providecommand \enquote  [1]{``#1''}%
\providecommand \bibnamefont  [1]{#1}%
\providecommand \bibfnamefont [1]{#1}%
\providecommand \citenamefont [1]{#1}%
\providecommand \href@noop [0]{\@secondoftwo}%
\providecommand \href [0]{\begingroup \@sanitize@url \@href}%
\providecommand \@href[1]{\@@startlink{#1}\@@href}%
\providecommand \@@href[1]{\endgroup#1\@@endlink}%
\providecommand \@sanitize@url [0]{\catcode `\\12\catcode `\$12\catcode
  `\&12\catcode `\#12\catcode `\^12\catcode `\_12\catcode `\%12\relax}%
\providecommand \@@startlink[1]{}%
\providecommand \@@endlink[0]{}%
\providecommand \url  [0]{\begingroup\@sanitize@url \@url }%
\providecommand \@url [1]{\endgroup\@href {#1}{\urlprefix }}%
\providecommand \urlprefix  [0]{URL }%
\providecommand \Eprint [0]{\href }%
\providecommand \doibase [0]{http://dx.doi.org/}%
\providecommand \selectlanguage [0]{\@gobble}%
\providecommand \bibinfo  [0]{\@secondoftwo}%
\providecommand \bibfield  [0]{\@secondoftwo}%
\providecommand \translation [1]{[#1]}%
\providecommand \BibitemOpen [0]{}%
\providecommand \bibitemStop [0]{}%
\providecommand \bibitemNoStop [0]{.\EOS\space}%
\providecommand \EOS [0]{\spacefactor3000\relax}%
\providecommand \BibitemShut  [1]{\csname bibitem#1\endcsname}%
\let\auto@bib@innerbib\@empty
\bibitem [{\citenamefont {Vincent}\ \emph {et~al.}(2012)\citenamefont {Vincent}
  \emph {et~al.}}]{Vincent2012}%
  \BibitemOpen
  \bibfield  {author} {\bibinfo {author} {\bibfnamefont {R.}~\bibnamefont
  {Vincent}} \emph {et~al.},\ }\href {\doibase 10.1038/nature11341} {\bibfield
  {journal} {\bibinfo  {journal} {Nature}\ }\textbf {\bibinfo {volume} {488}},\
  \bibinfo {pages} {357} (\bibinfo {year} {2012})}\BibitemShut {NoStop}%
\bibitem [{\citenamefont {Bogani}\ and\ \citenamefont
  {Wernsdorfer}(2008)}]{Bogani2008}%
  \BibitemOpen
  \bibfield  {author} {\bibinfo {author} {\bibfnamefont {L.}~\bibnamefont
  {Bogani}}\ and\ \bibinfo {author} {\bibfnamefont {W.}~\bibnamefont
  {Wernsdorfer}},\ }\href {\doibase 10.1038/nmat2133} {\bibfield  {journal}
  {\bibinfo  {journal} {Nat. Mater.}\ }\textbf {\bibinfo {volume} {7}},\
  \bibinfo {pages} {179} (\bibinfo {year} {2008})}\BibitemShut {NoStop}%
\bibitem [{\citenamefont {Leuenberger}\ and\ \citenamefont
  {Loss}(2001)}]{Leuenberger2001}%
  \BibitemOpen
  \bibfield  {author} {\bibinfo {author} {\bibfnamefont {M.~N.}\ \bibnamefont
  {Leuenberger}}\ and\ \bibinfo {author} {\bibfnamefont {D.}~\bibnamefont
  {Loss}},\ }\href {\doibase 10.1038/35071024} {\bibfield  {journal} {\bibinfo
  {journal} {Nature}\ }\textbf {\bibinfo {volume} {410}},\ \bibinfo {pages}
  {789} (\bibinfo {year} {2001})}\BibitemShut {NoStop}%
\bibitem [{\citenamefont {Gatteschi}\ and\ \citenamefont
  {Sessoli}(2003)}]{Gatteschi2003}%
  \BibitemOpen
  \bibfield  {author} {\bibinfo {author} {\bibfnamefont {D.}~\bibnamefont
  {Gatteschi}}\ and\ \bibinfo {author} {\bibfnamefont {R.}~\bibnamefont
  {Sessoli}},\ }\href {\doibase 10.1002/anie.200390099} {\bibfield  {journal}
  {\bibinfo  {journal} {Angew. Chemie Int. Ed.}\ }\textbf {\bibinfo {volume}
  {42}},\ \bibinfo {pages} {268} (\bibinfo {year} {2003})}\BibitemShut
  {NoStop}%
\bibitem [{\citenamefont {Sessoli}\ \emph {et~al.}(1993)\citenamefont
  {Sessoli}, \citenamefont {Gatteschi}, \citenamefont {Caneschi},\ and\
  \citenamefont {Novak}}]{Sessoli1993}%
  \BibitemOpen
  \bibfield  {author} {\bibinfo {author} {\bibfnamefont {R.}~\bibnamefont
  {Sessoli}}, \bibinfo {author} {\bibfnamefont {D.}~\bibnamefont {Gatteschi}},
  \bibinfo {author} {\bibfnamefont {a.}~\bibnamefont {Caneschi}}, \ and\
  \bibinfo {author} {\bibfnamefont {M.~a.}\ \bibnamefont {Novak}},\ }\href
  {\doibase 10.1038/365141a0} {\bibfield  {journal} {\bibinfo  {journal}
  {Nature}\ }\textbf {\bibinfo {volume} {365}},\ \bibinfo {pages} {141}
  (\bibinfo {year} {1993})}\BibitemShut {NoStop}%
\bibitem [{\citenamefont {Casimir}\ and\ \citenamefont
  {du~Pr{\'{e}}}(1938)}]{casimir1938a}%
  \BibitemOpen
  \bibfield  {author} {\bibinfo {author} {\bibfnamefont {H.}~\bibnamefont
  {Casimir}}\ and\ \bibinfo {author} {\bibfnamefont {F.}~\bibnamefont
  {du~Pr{\'{e}}}},\ }\href {\doibase 10.1016/S0031-8914(38)80164-6} {\bibfield
  {journal} {\bibinfo  {journal} {Physica}\ }\textbf {\bibinfo {volume} {5}},\
  \bibinfo {pages} {507} (\bibinfo {year} {1938})}\BibitemShut {NoStop}%
\bibitem [{\citenamefont {McConnell}(1980)}]{mcconnell1980}%
  \BibitemOpen
  \bibfield  {author} {\bibinfo {author} {\bibfnamefont {J.~R.}\ \bibnamefont
  {McConnell}},\ }\href
  {https://books.google.com/books?id=ehQvAAAAIAAJ{\&}pgis=1} {\emph {\bibinfo
  {title} {{Rotational Brownian motion and dielectric theory}}}}\ (\bibinfo
  {publisher} {Academic Press},\ \bibinfo {year} {1980})\BibitemShut {NoStop}%
\bibitem [{\citenamefont {Cole}\ and\ \citenamefont {Cole}(1941)}]{cole1941}%
  \BibitemOpen
  \bibfield  {author} {\bibinfo {author} {\bibfnamefont {K.~S.}\ \bibnamefont
  {Cole}}\ and\ \bibinfo {author} {\bibfnamefont {R.~H.}\ \bibnamefont
  {Cole}},\ }\href {\doibase 10.1063/1.1750906} {\bibfield  {journal} {\bibinfo
   {journal} {J. Chem. Phys.}\ }\textbf {\bibinfo {volume} {9}},\ \bibinfo
  {pages} {341} (\bibinfo {year} {1941})}\BibitemShut {NoStop}%
\bibitem [{\citenamefont {Ishikawa}\ \emph {et~al.}(2003)\citenamefont
  {Ishikawa} \emph {et~al.}}]{Ishikawa2003}%
  \BibitemOpen
  \bibfield  {author} {\bibinfo {author} {\bibfnamefont {N.}~\bibnamefont
  {Ishikawa}} \emph {et~al.},\ }\href {\doibase 10.1021/ja029629n} {\bibfield
  {journal} {\bibinfo  {journal} {J. Am. Chem. Soc.}\ }\textbf {\bibinfo
  {volume} {125}},\ \bibinfo {pages} {8694} (\bibinfo {year}
  {2003})}\BibitemShut {NoStop}%
\bibitem [{\citenamefont {Sessoli}\ and\ \citenamefont
  {Powell}(2009)}]{Sessoli2009}%
  \BibitemOpen
  \bibfield  {author} {\bibinfo {author} {\bibfnamefont {R.}~\bibnamefont
  {Sessoli}}\ and\ \bibinfo {author} {\bibfnamefont {A.~K.}\ \bibnamefont
  {Powell}},\ }\href {\doibase 10.1016/j.ccr.2008.12.014} {\bibfield  {journal}
  {\bibinfo  {journal} {Coord. Chem. Rev.}\ }\textbf {\bibinfo {volume}
  {253}},\ \bibinfo {pages} {2328} (\bibinfo {year} {2009})}\BibitemShut
  {NoStop}%
\bibitem [{\citenamefont {Woodruff}\ \emph {et~al.}(2013)\citenamefont
  {Woodruff}, \citenamefont {Winpenny},\ and\ \citenamefont
  {Layfield}}]{Woodruff2013}%
  \BibitemOpen
  \bibfield  {author} {\bibinfo {author} {\bibfnamefont {D.~N.}\ \bibnamefont
  {Woodruff}}, \bibinfo {author} {\bibfnamefont {R.~E.~P.}\ \bibnamefont
  {Winpenny}}, \ and\ \bibinfo {author} {\bibfnamefont {R.~A.}\ \bibnamefont
  {Layfield}},\ }\href {\doibase 10.1021/cr400018q} {\bibfield  {journal}
  {\bibinfo  {journal} {Chem. Rev.}\ }\textbf {\bibinfo {volume} {113}},\
  \bibinfo {pages} {5110} (\bibinfo {year} {2013})}\BibitemShut {NoStop}%
\bibitem [{\citenamefont {AlDamen}\ \emph {et~al.}(2009)\citenamefont {AlDamen}
  \emph {et~al.}}]{AlDamen2009}%
  \BibitemOpen
  \bibfield  {author} {\bibinfo {author} {\bibfnamefont {M.~A.}\ \bibnamefont
  {AlDamen}} \emph {et~al.},\ }\href {\doibase 10.1021/ic801630z} {\bibfield
  {journal} {\bibinfo  {journal} {Inorg. Chem.}\ }\textbf {\bibinfo {volume}
  {48}},\ \bibinfo {pages} {3467} (\bibinfo {year} {2009})}\BibitemShut
  {NoStop}%
\bibitem [{\citenamefont {Blagg}\ \emph {et~al.}(2013)\citenamefont {Blagg}
  \emph {et~al.}}]{Blagg2013b}%
  \BibitemOpen
  \bibfield  {author} {\bibinfo {author} {\bibnamefont {Blagg}} \emph
  {et~al.},\ }\href {\doibase 10.1038/nchem.1707} {\bibfield  {journal}
  {\bibinfo  {journal} {Nat. Chem.}\ }\textbf {\bibinfo {volume} {5}},\
  \bibinfo {pages} {673} (\bibinfo {year} {2013})}\BibitemShut {NoStop}%
\bibitem [{\citenamefont {Hewitt}\ \emph {et~al.}(2010)\citenamefont {Hewitt}
  \emph {et~al.}}]{Hewitt2010}%
  \BibitemOpen
  \bibfield  {author} {\bibinfo {author} {\bibfnamefont {I.~J.}\ \bibnamefont
  {Hewitt}} \emph {et~al.},\ }\href {\doibase 10.1002/anie.201002691}
  {\bibfield  {journal} {\bibinfo  {journal} {Angew. Chemie Int. Ed.}\ }\textbf
  {\bibinfo {volume} {49}},\ \bibinfo {pages} {6352} (\bibinfo {year}
  {2010})}\BibitemShut {NoStop}%
\bibitem [{\citenamefont {Lin}\ \emph {et~al.}(2009)\citenamefont {Lin} \emph
  {et~al.}}]{Lin2009}%
  \BibitemOpen
  \bibfield  {author} {\bibinfo {author} {\bibfnamefont {P.-H.}\ \bibnamefont
  {Lin}} \emph {et~al.},\ }\href {\doibase 10.1002/anie.200903199} {\bibfield
  {journal} {\bibinfo  {journal} {Angew. Chemie Int. Ed.}\ }\textbf {\bibinfo
  {volume} {48}},\ \bibinfo {pages} {9489} (\bibinfo {year}
  {2009})}\BibitemShut {NoStop}%
\bibitem [{\citenamefont {Guo}\ \emph {et~al.}(2011{\natexlab{a}})\citenamefont
  {Guo} \emph {et~al.}}]{Guo2011c}%
  \BibitemOpen
  \bibfield  {author} {\bibinfo {author} {\bibfnamefont {F.~S.}\ \bibnamefont
  {Guo}} \emph {et~al.},\ }\href {\doibase 10.1002/chem.201002296} {\bibfield
  {journal} {\bibinfo  {journal} {Chem. - A Eur. J.}\ }\textbf {\bibinfo
  {volume} {17}},\ \bibinfo {pages} {2458} (\bibinfo {year}
  {2011}{\natexlab{a}})}\BibitemShut {NoStop}%
\bibitem [{\citenamefont {Amjad}\ \emph {et~al.}(2016)\citenamefont {Amjad},
  \citenamefont {Figuerola}, \citenamefont {Caneschi},\ and\ \citenamefont
  {Sorace}}]{Amjad2016}%
  \BibitemOpen
  \bibfield  {author} {\bibinfo {author} {\bibfnamefont {A.}~\bibnamefont
  {Amjad}}, \bibinfo {author} {\bibfnamefont {A.}~\bibnamefont {Figuerola}},
  \bibinfo {author} {\bibfnamefont {A.}~\bibnamefont {Caneschi}}, \ and\
  \bibinfo {author} {\bibfnamefont {L.}~\bibnamefont {Sorace}},\ }\href
  {\doibase 10.3390/magnetochemistry2020027} {\bibfield  {journal} {\bibinfo
  {journal} {Magnetochemistry}\ }\textbf {\bibinfo {volume} {2}},\ \bibinfo
  {pages} {27} (\bibinfo {year} {2016})}\BibitemShut {NoStop}%
\bibitem [{\citenamefont {Guo}\ \emph {et~al.}(2011{\natexlab{b}})\citenamefont
  {Guo} \emph {et~al.}}]{Guo2011a}%
  \BibitemOpen
  \bibfield  {author} {\bibinfo {author} {\bibfnamefont {Y.-N.~Y.}\
  \bibnamefont {Guo}} \emph {et~al.},\ }\href {\doibase 10.1021/ja205035g}
  {\bibfield  {journal} {\bibinfo  {journal} {J. Am. Chem. Soc.}\ }\textbf
  {\bibinfo {volume} {133}},\ \bibinfo {pages} {11948} (\bibinfo {year}
  {2011}{\natexlab{b}})}\BibitemShut {NoStop}%
\bibitem [{\citenamefont {Guo}\ \emph {et~al.}(2010)\citenamefont {Guo} \emph
  {et~al.}}]{Guo2010}%
  \BibitemOpen
  \bibfield  {author} {\bibinfo {author} {\bibfnamefont {Y.-N.}\ \bibnamefont
  {Guo}} \emph {et~al.},\ }\href {\doibase 10.1021/ja103018m} {\bibfield
  {journal} {\bibinfo  {journal} {J. Am. Chem. Soc.}\ }\textbf {\bibinfo
  {volume} {132}},\ \bibinfo {pages} {8538} (\bibinfo {year}
  {2010})}\BibitemShut {NoStop}%
\bibitem [{\citenamefont {Hewitt}\ \emph {et~al.}(2009)\citenamefont {Hewitt}
  \emph {et~al.}}]{Hewitt2009}%
  \BibitemOpen
  \bibfield  {author} {\bibinfo {author} {\bibnamefont {Hewitt}} \emph
  {et~al.},\ }\href {\doibase 10.1039/b908194a} {\bibfield  {journal} {\bibinfo
   {journal} {Chem. Commun. (Camb).}\ }\textbf {\bibinfo {volume} {3}},\
  \bibinfo {pages} {6765} (\bibinfo {year} {2009})}\BibitemShut {NoStop}%
\bibitem [{\citenamefont {Rinehart}\ \emph {et~al.}(2010)\citenamefont
  {Rinehart}, \citenamefont {Meihaus},\ and\ \citenamefont
  {Long}}]{Rinehart2010}%
  \BibitemOpen
  \bibfield  {author} {\bibinfo {author} {\bibfnamefont {J.~D.}\ \bibnamefont
  {Rinehart}}, \bibinfo {author} {\bibfnamefont {K.~R.}\ \bibnamefont
  {Meihaus}}, \ and\ \bibinfo {author} {\bibfnamefont {J.~R.}\ \bibnamefont
  {Long}},\ }\href {\doibase 10.1021/ja1009019} {\bibfield  {journal} {\bibinfo
   {journal} {J. Am. Chem. Soc.}\ }\textbf {\bibinfo {volume} {132}},\ \bibinfo
  {pages} {7572} (\bibinfo {year} {2010})}\BibitemShut {NoStop}%
\bibitem [{\citenamefont {Miklovi{\v{c}}}\ \emph {et~al.}(2015)\citenamefont
  {Miklovi{\v{c}}}, \citenamefont {Valigura}, \citenamefont {Bo{\v{c}}a},\ and\
  \citenamefont {Titi{\v{s}}}}]{Miklovic2015}%
  \BibitemOpen
  \bibfield  {author} {\bibinfo {author} {\bibfnamefont {J.}~\bibnamefont
  {Miklovi{\v{c}}}}, \bibinfo {author} {\bibfnamefont {D.}~\bibnamefont
  {Valigura}}, \bibinfo {author} {\bibfnamefont {R.}~\bibnamefont
  {Bo{\v{c}}a}}, \ and\ \bibinfo {author} {\bibfnamefont {J.}~\bibnamefont
  {Titi{\v{s}}}},\ }\href {\doibase 10.1039/C5DT01213A} {\bibfield  {journal}
  {\bibinfo  {journal} {Dalton Trans.}\ }\textbf {\bibinfo {volume} {44}},\
  \bibinfo {pages} {12484} (\bibinfo {year} {2015})}\BibitemShut {NoStop}%
\bibitem [{\citenamefont {Holmberg}\ \emph {et~al.}(2015)\citenamefont
  {Holmberg} \emph {et~al.}}]{Holmberg2015}%
  \BibitemOpen
  \bibfield  {author} {\bibinfo {author} {\bibfnamefont {R.~J.}\ \bibnamefont
  {Holmberg}} \emph {et~al.},\ }\href {\doibase 10.1039/C5DT04072H} {\bibfield
  {journal} {\bibinfo  {journal} {Dalton Trans.}\ }\textbf {\bibinfo {volume}
  {44}},\ \bibinfo {pages} {20321} (\bibinfo {year} {2015})}\BibitemShut
  {NoStop}%
\bibitem [{\citenamefont {Gupta}\ \emph {et~al.}(2016)\citenamefont {Gupta},
  \citenamefont {Rajeshkumar}, \citenamefont {Rajaraman},\ and\ \citenamefont
  {Murugavel}}]{Gupta2016}%
  \BibitemOpen
  \bibfield  {author} {\bibinfo {author} {\bibfnamefont {S.~K.}\ \bibnamefont
  {Gupta}}, \bibinfo {author} {\bibfnamefont {T.}~\bibnamefont {Rajeshkumar}},
  \bibinfo {author} {\bibfnamefont {G.}~\bibnamefont {Rajaraman}}, \ and\
  \bibinfo {author} {\bibfnamefont {R.}~\bibnamefont {Murugavel}},\ }\href
  {\doibase 10.1039/C6CC03066A} {\bibfield  {journal} {\bibinfo  {journal}
  {Chem. Commun.}\ }\textbf {\bibinfo {volume} {52}},\ \bibinfo {pages} {7168}
  (\bibinfo {year} {2016})}\BibitemShut {NoStop}%
\bibitem [{\citenamefont {Jeletic}\ \emph {et~al.}(2011)\citenamefont {Jeletic}
  \emph {et~al.}}]{Jeletic2011}%
  \BibitemOpen
  \bibfield  {author} {\bibinfo {author} {\bibnamefont {Jeletic}} \emph
  {et~al.},\ }\href {\doibase 10.1021/ja207891y} {\bibfield  {journal}
  {\bibinfo  {journal} {J. Am. Chem. Soc.}\ }\textbf {\bibinfo {volume}
  {133}},\ \bibinfo {pages} {19286} (\bibinfo {year} {2011})}\BibitemShut
  {NoStop}%
\bibitem [{\citenamefont {Habib}\ \emph {et~al.}(2013)\citenamefont {Habib}
  \emph {et~al.}}]{Habib2013}%
  \BibitemOpen
  \bibfield  {author} {\bibinfo {author} {\bibfnamefont {F.}~\bibnamefont
  {Habib}} \emph {et~al.},\ }\href {\doibase 10.1002/anie.201303005} {\bibfield
   {journal} {\bibinfo  {journal} {Angew. Chemie Int. Ed.}\ }\textbf {\bibinfo
  {volume} {52}},\ \bibinfo {pages} {11290} (\bibinfo {year}
  {2013})}\BibitemShut {NoStop}%
\bibitem [{\citenamefont {Habib}\ \emph {et~al.}(2015)\citenamefont {Habib},
  \citenamefont {Korobkov},\ and\ \citenamefont {Murugesu}}]{Habib2015}%
  \BibitemOpen
  \bibfield  {author} {\bibinfo {author} {\bibfnamefont {F.}~\bibnamefont
  {Habib}}, \bibinfo {author} {\bibfnamefont {I.}~\bibnamefont {Korobkov}}, \
  and\ \bibinfo {author} {\bibfnamefont {M.}~\bibnamefont {Murugesu}},\ }\href
  {\doibase 10.1039/c5dt00258c} {\bibfield  {journal} {\bibinfo  {journal}
  {Dalton Trans.}\ }\textbf {\bibinfo {volume} {44}},\ \bibinfo {pages} {6368}
  (\bibinfo {year} {2015})}\BibitemShut {NoStop}%
\bibitem [{\citenamefont {Ruiz}\ \emph {et~al.}(2012)\citenamefont {Ruiz} \emph
  {et~al.}}]{Ruiz2012}%
  \BibitemOpen
  \bibfield  {author} {\bibinfo {author} {\bibfnamefont {J.}~\bibnamefont
  {Ruiz}} \emph {et~al.},\ }\href {\doibase 10.1039/c2cc32518g} {\bibfield
  {journal} {\bibinfo  {journal} {Chem. Commun. (Camb).}\ }\textbf {\bibinfo
  {volume} {48}},\ \bibinfo {pages} {7916} (\bibinfo {year}
  {2012})}\BibitemShut {NoStop}%
\bibitem [{\citenamefont {Lucaccini}\ \emph {et~al.}(2016)\citenamefont
  {Lucaccini} \emph {et~al.}}]{Lucaccini2016b}%
  \BibitemOpen
  \bibfield  {author} {\bibinfo {author} {\bibfnamefont {E.}~\bibnamefont
  {Lucaccini}} \emph {et~al.},\ }\href {\doibase 10.1002/chem.201505211}
  {\bibfield  {journal} {\bibinfo  {journal} {Chem. - A Eur. J.}\ }\textbf
  {\bibinfo {volume} {22}},\ \bibinfo {pages} {5552} (\bibinfo {year}
  {2016})}\BibitemShut {NoStop}%
\bibitem [{\citenamefont {Gregson}\ \emph {et~al.}(2016)\citenamefont {Gregson}
  \emph {et~al.}}]{Gregson2015}%
  \BibitemOpen
  \bibfield  {author} {\bibinfo {author} {\bibfnamefont {M.}~\bibnamefont
  {Gregson}} \emph {et~al.},\ }\href {\doibase 10.1039/C5SC03111G} {\bibfield
  {journal} {\bibinfo  {journal} {Chem. Sci.}\ }\textbf {\bibinfo {volume}
  {7}},\ \bibinfo {pages} {155} (\bibinfo {year} {2016})}\BibitemShut {NoStop}%
\bibitem [{\citenamefont {Li}\ \emph {et~al.}(2016)\citenamefont {Li} \emph
  {et~al.}}]{Li2016}%
  \BibitemOpen
  \bibfield  {author} {\bibinfo {author} {\bibfnamefont {J.}~\bibnamefont {Li}}
  \emph {et~al.},\ }\href {\doibase 10.1039/C6DT00979D} {\bibfield  {journal}
  {\bibinfo  {journal} {Dalton Trans.}\ }\textbf {\bibinfo {volume} {45}},\
  \bibinfo {pages} {9279} (\bibinfo {year} {2016})}\BibitemShut {NoStop}%
\bibitem [{\citenamefont {Cosquer}\ \emph {et~al.}(2013)\citenamefont {Cosquer}
  \emph {et~al.}}]{Cosquer2013}%
  \BibitemOpen
  \bibfield  {author} {\bibinfo {author} {\bibfnamefont {G.}~\bibnamefont
  {Cosquer}} \emph {et~al.},\ }\href {\doibase 10.1002/chem.201300397}
  {\bibfield  {journal} {\bibinfo  {journal} {Chem. - A Eur. J.}\ }\textbf
  {\bibinfo {volume} {19}},\ \bibinfo {pages} {7895} (\bibinfo {year}
  {2013})}\BibitemShut {NoStop}%
\bibitem [{Note1()}]{Note1}%
  \BibitemOpen
  \bibinfo {note} {The studies of diluted U(H$_2$BPz$_2$)$_3$ have shown a
  relationship between the existence of the secondary relaxation process and
  the degree of dilution, pointing on the importance of intermolecular
  interaction for its observation \protect \cite
  {Meihaus2011,*Meihaus2015}}\BibitemShut {NoStop}%
\bibitem [{\citenamefont {Gatteschi}\ \emph {et~al.}(2006)\citenamefont
  {Gatteschi}, \citenamefont {Sessoli},\ and\ \citenamefont
  {Villain}}]{Gatteschi2006}%
  \BibitemOpen
  \bibfield  {author} {\bibinfo {author} {\bibfnamefont {D.}~\bibnamefont
  {Gatteschi}}, \bibinfo {author} {\bibfnamefont {R.}~\bibnamefont {Sessoli}},
  \ and\ \bibinfo {author} {\bibfnamefont {J.}~\bibnamefont {Villain}},\ }\href
  {\doibase 10.1093/acprof:oso/9780198567530.001.0001} {\emph {\bibinfo {title}
  {{Molecular Nanomagnets}}}}\ (\bibinfo  {publisher} {Oxford University
  Press},\ \bibinfo {year} {2006})\BibitemShut {NoStop}%
\bibitem [{\citenamefont {Garanin}(2012)}]{garanin2008}%
  \BibitemOpen
  \bibfield  {author} {\bibinfo {author} {\bibfnamefont {D.~A.}\ \bibnamefont
  {Garanin}},\ }\href {\doibase 10.1002/9781118135242.ch4} {\bibfield
  {journal} {\bibinfo  {journal} {Adv. Chem. Phys.}\ }\textbf {\bibinfo
  {volume} {147}},\ \bibinfo {pages} {213} (\bibinfo {year} {2012})},\ \Eprint
  {http://arxiv.org/abs/0805.0391} {arXiv:0805.0391} \BibitemShut {NoStop}%
\bibitem [{\citenamefont {Leuenberger}\ and\ \citenamefont
  {Loss}(2000)}]{Leuenberger2000}%
  \BibitemOpen
  \bibfield  {author} {\bibinfo {author} {\bibfnamefont {M.}~\bibnamefont
  {Leuenberger}}\ and\ \bibinfo {author} {\bibfnamefont {D.}~\bibnamefont
  {Loss}},\ }\href {http://prb.aps.org/abstract/PRB/v61/i2/p1286{\_}1}
  {\bibfield  {journal} {\bibinfo  {journal} {Phys. Rev. B}\ }\textbf {\bibinfo
  {volume} {61}},\ \bibinfo {pages} {1286} (\bibinfo {year}
  {2000})}\BibitemShut {NoStop}%
\bibitem [{\citenamefont {Abragam}\ and\ \citenamefont
  {Bleaney}(1970)}]{abragam1970}%
  \BibitemOpen
  \bibfield  {author} {\bibinfo {author} {\bibfnamefont {A.}~\bibnamefont
  {Abragam}}\ and\ \bibinfo {author} {\bibfnamefont {B.}~\bibnamefont
  {Bleaney}},\ }\href@noop {} {\emph {\bibinfo {title} {{EPR of Transition
  Ions}}}}\ (\bibinfo  {publisher} {Clarendon Press, Oxford},\ \bibinfo {year}
  {1970})\BibitemShut {NoStop}%
\bibitem [{Note2()}]{Note2}%
  \BibitemOpen
  \bibinfo {note} {Generally, a tunneling relaxation rate, $\Gamma _{\protect
  \mathrm {tunneling}}$, should be added to the right hand side of this
  expression, however, it is suppressed in the present case}\BibitemShut
  {NoStop}%
\bibitem [{a1()}]{a1}%
  \BibitemOpen
  \href@noop {} {\bibinfo  {journal} {L. T. A. Ho and L. F. Chibotaru (to be
  published)}\ }\BibitemShut {NoStop}%
\bibitem [{Note3()}]{Note3}%
  \BibitemOpen
\bibfield  {journal} {  }\bibinfo {note} {Given the explicit involvement of the
  state $\mathinner {|{3}\delimiter "526930B }$ in the present description of
  relaxation, Eq. (\ref {eq:general linear response}), $\Gamma _{31}$ refers to
  a direct relaxation process. However, we still call it $\Gamma _{\protect
  \mathrm {Orbach}}$ for the sake of convenience}\BibitemShut {NoStop}%
\bibitem [{\citenamefont {Guo}\ \emph {et~al.}(2011{\natexlab{c}})\citenamefont
  {Guo}, \citenamefont {Xu}, \citenamefont {Guo},\ and\ \citenamefont
  {Tang}}]{Guo2011b}%
  \BibitemOpen
  \bibfield  {author} {\bibinfo {author} {\bibfnamefont {Y.-N.}\ \bibnamefont
  {Guo}}, \bibinfo {author} {\bibfnamefont {G.-F.}\ \bibnamefont {Xu}},
  \bibinfo {author} {\bibfnamefont {Y.}~\bibnamefont {Guo}}, \ and\ \bibinfo
  {author} {\bibfnamefont {J.}~\bibnamefont {Tang}},\ }\href {\doibase
  10.1039/c1dt10474h} {\bibfield  {journal} {\bibinfo  {journal} {Dalton
  Trans.}\ }\textbf {\bibinfo {volume} {40}},\ \bibinfo {pages} {9953}
  (\bibinfo {year} {2011}{\natexlab{c}})}\BibitemShut {NoStop}%
\bibitem [{\citenamefont {Meihaus}\ \emph {et~al.}(2011)\citenamefont
  {Meihaus}, \citenamefont {Rinehart},\ and\ \citenamefont
  {Long}}]{Meihaus2011}%
  \BibitemOpen
  \bibfield  {author} {\bibinfo {author} {\bibfnamefont {K.~R.}\ \bibnamefont
  {Meihaus}}, \bibinfo {author} {\bibfnamefont {J.~D.}\ \bibnamefont
  {Rinehart}}, \ and\ \bibinfo {author} {\bibfnamefont {J.~R.}\ \bibnamefont
  {Long}},\ }\href@noop {} {\bibfield  {journal} {\bibinfo  {journal}
  {Inorganic chemistry}\ }\textbf {\bibinfo {volume} {50}},\ \bibinfo {pages}
  {8484} (\bibinfo {year} {2011})}\BibitemShut {NoStop}%
\bibitem [{\citenamefont {Meihaus}\ and\ \citenamefont
  {Long}(2015)}]{Meihaus2015}%
  \BibitemOpen
  \bibfield  {author} {\bibinfo {author} {\bibfnamefont {K.~R.}\ \bibnamefont
  {Meihaus}}\ and\ \bibinfo {author} {\bibfnamefont {J.~R.}\ \bibnamefont
  {Long}},\ }\href@noop {} {\bibfield  {journal} {\bibinfo  {journal} {Dalton
  Transactions}\ }\textbf {\bibinfo {volume} {44}},\ \bibinfo {pages} {2517}
  (\bibinfo {year} {2015})}\BibitemShut {NoStop}%
\end{thebibliography}
%

\end{document}